\documentclass[reprint, prb, superscriptaddress, nofootinbib]{revtex4-2}
\usepackage{amsmath,amssymb}
\usepackage{stmaryrd}
\usepackage{array}
\usepackage{latexsym}
\usepackage{CJK}
\usepackage{graphicx}
\usepackage{indentfirst}
\usepackage{listings}
\usepackage{xcolor}
\usepackage{booktabs}
\usepackage{stmaryrd}
\usepackage{multirow}
\usepackage{verbatim}
\usepackage{makecell}
\usepackage{lineno,hyperref}
\usepackage{longtable}
\usepackage{hyperref}
\usepackage{upgreek}
\usepackage{mathtools}
\hypersetup{colorlinks,breaklinks,
            linkcolor=blue,urlcolor=blue,anchorcolor=blue,citecolor=blue}

\newcommand{\rmd}{\mathrm{d}}

\begin{document}

\title{Why Grain Growth is Not Curvature Flow}

\author{Caihao Qiu}
\affiliation{Department of Materials Science and Engineering, City University of Hong Kong,  Hong Kong SAR, China}
\author{David J. Srolovitz}
\affiliation{Department of Mechanical Engineering, The University of Hong Kong, Pokfulam Road, Hong Kong SAR, China}
\affiliation{Materials Innovation Institute for Life Sciences and Energy (MILES), The University of Hong Kong, Shenzhen, China}
\author{Gregory S. Rohrer}
\affiliation{Department of Materials Science and Engineering, Carnegie Mellon University, Pittsburgh, PA, USA}
\author{Jian Han}
\email{jianhan@cityu.edu.hk}
\affiliation{Department of Materials Science and Engineering, City University of Hong Kong, Hong Kong SAR, China}
\author{Marco Salvalaglio}
\email{marco.salvalaglio@tu-dresden.de}
\affiliation{Institute of Scientific Computing, TU Dresden, 01062 Dresden, Germany}
\affiliation{Dresden Center for Computational Materials Science, TU Dresden, 01062 Dresden, Germany}

\date{\today}

\begin{abstract}
Grain growth in polycrystals is traditionally considered a capillarity-driven process, 
where grain boundaries (GBs) migrate toward their centers of curvature (i.e., mean curvature flow) with a velocity proportional to the local curvature (including extensions to account for  anisotropic GB energy and mobility).
Experimental and simulation evidence shows that this simplistic view is untrue.
We demonstrate that the failure of the classical mean curvature flow description of grain growth mainly originates from the shear deformation naturally coupled with GB motion (i.e., shear coupling).
Our findings are built on large-scale microstructure evolution simulations incorporating the fundamental (crystallography-respecting) microscopic mechanism of GB migration.
The nature of the deviations from curvature flow revealed in our simulations is consistent with observations in recent experimental studies on different materials.
This work also demonstrates how to incorporate the mechanical effects that are essential to the accurate prediction of microstructure evolution.
\end{abstract}

\maketitle

Polycrystalline materials consist of numerous single crystals (grains) with different crystallographic orientations separated by interfaces/grain boundaries (GBs).
Polycrystals are ubiquitous in both natural and man-made materials; most engineering materials are polycrystalline.
The microstructure, namely the geometry of the grain ensemble and GB network, plays a major role in establishing the (physical, mechanical,...) properties of polycrystalline materials.
The classical theory of microstructure evolution focuses on the relaxation of the excess energy of a GB network by GB migration, resulting in grain annihilation.
GBs tend to migrate in the direction that reduces their area, that is towards their center of curvature.
This gives rise to the well-known description of grain growth as mean curvature ($\kappa$) flow;  the GB velocity may be expressed as $v = M \gamma \kappa$, where $M$ is the GB mobility and $\gamma$ is the (isotropic) GB energy \cite{smith1948grains,von1952metal,mullins1956two,macpherson2007vonneumann}.
Although this model describes the evolution of similar-looking microstructures, such as relatively dry soap froths, it neglects the fundamental nature of the material in the grains (crystals rather than gases).  
The crystalline nature of the grains implies that GB properties are anisotropic. 
This is often accounted for by extending the classical mean curvature flow description to the case of anisotropic GB mobility and/or energy depending on GB macroscopic degrees of freedom; i.e.,  weighted mean curvature flow~\cite{taylor1992overview,kazaryan2002grain}.
Importantly, the material itself also supports mechanical deformation, i.e., stress and strain.

Several recent investigations clearly show that a (weighted) mean curvature flow description of grain growth fails to explain experimental and atomistic simulation observations.
Phenomena that point to this failure include the existence of stress-driven grain growth \cite{legros2008situ,mompiou2009grain,sharon2011stress}, grain rotation \cite{harris1998grain,farkas2007linear, tian2024grain}, GB sliding \cite{schafer2012competing, wang2022tracking}, grain growth acceleration under oscillatory fields~\cite{qiu2024brownian}, and peculiar grain-size distributions \cite{barmak2013grain,backofen2014capturing}.
Recent experimental measurements and molecular dynamics (MD) simulations found that, in some systems, GB velocity is \emph{not} correlated with GB mean curvature \cite{bhattacharya2021grain,xu2024grain,bizana2023kinetics}.
Moreover, many observations show that GB migration is accompanied by shear across the GB, both in bicrystals and polycrystals; i.e., shear-coupled migration associated with defect (disconnection) motion along GBs \cite{li1953stress,cahn2006coupling,han2018grain,qiu2024brownian}.

\begin{figure*}[t]
  \centering
\includegraphics[width=0.99\linewidth]{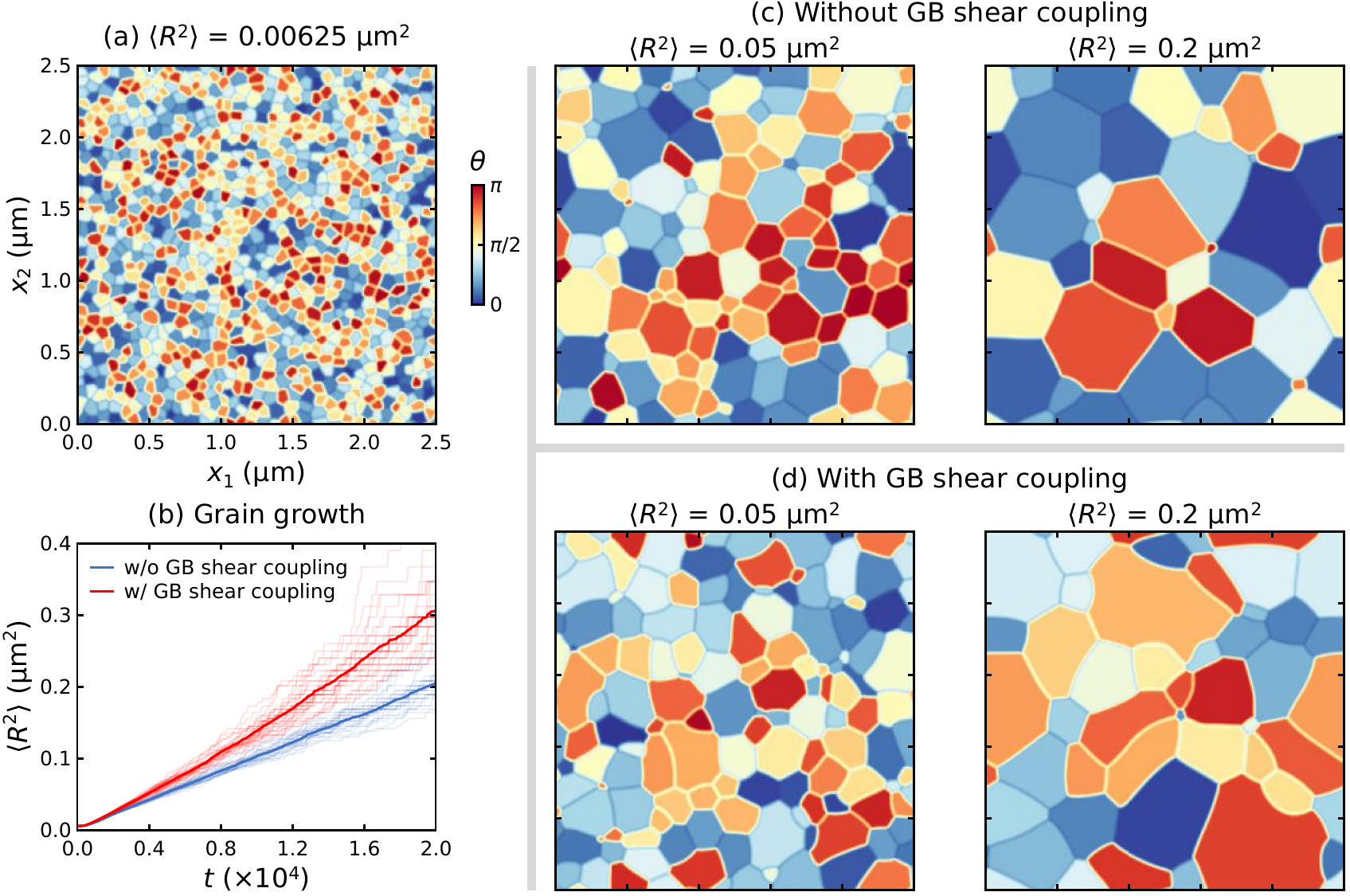}\hspace{-1.78em}%
\caption{\textbf{2D grain growth simulations.} (a) Initial microstructure and orientation $\theta$ field of a typical 1000 grain polycrystal.
(b) Mean grain size $\langle R^2 \rangle$ versus time $t$ during grain growth.
Each solid and thick curve is averaged over 30 independent PF simulations (i.e., the light and thin curves)  with and without GB shear coupling.
Microstructure evolution (c) without  GB shear coupling (mean curvature flow) and (d) with shear coupling (shear modulus is that of Al).
}
\label{fig_poly}
\end{figure*}

Shear-coupled migration plays a key role in accounting for much of the phenomenology associated with GB motion.
The coupling factor, $\beta = v_\parallel / v_\perp$, relates the GB migration rate  $v_\perp$ and the GB shear rate $v_\parallel$.
When shear-coupled GB migration is constrained (e.g., by triple junctions), an internal stress field is generated \cite{han2018grain, thomas2017reconciling}.
The stress, in turn, affects GB migration and influences grain growth \cite{thomas2017reconciling}.
We recently unified these elastic and curvature effects in a bicrystallography-respecting continuum GB migration model \cite{zhang2017equation,han2022disconnection} and demonstrated that GB shear coupling naturally leads to internal stress generation \cite{han2022disconnection,sal2022disconnection}, GB faceting transition \cite{qiu2023interface}, and grain rotation \cite{qiu2024disconnection} during GB migration in bicrystal geometries.
Nonetheless, two critical questions remain: (1) how does shear-coupling influence large-scale microstructure evolution? and (2) can shear-coupling account for the recent experimental observations that grain growth is not curvature flow?
Fortunately, both simulation and experimental observations have reached a level where detailed comparisons may now be made at the microscopic and statistical levels to quantitively answer these questions.

In this work, we demonstrate that the internal stresses generated by shear-coupled GB motion is the main cause for the poor correlations between the GB velocity and the curvature during grain growth in polycrystals.
This is achieved through a series of large-scale grain growth simulations.
They are performed using the bicrystallography-respecting continuum model for GB migration mentioned above, implemented in a diffuse-interface approach \cite{han2022disconnection,sal2022disconnection} extending the classical phase field (PF) model of grain growth \cite{Chen2002,Steinbach_2009}.
Simulated microstructure evolution, accounting for both GB curvature effect and internal stress, reproduces the scatter in the curvature-velocity relationship reported in several recent experimental and atomistic simulation studies.
We also demonstrate that other proposed explanations for the failure of the application of curvature flow to grain growth are insufficient to account for the recent experimental observations.

\section{Results}

\subsection{Effect of internal stress on 2D grain growth}

We first discuss a series of PF simulations on grain growth of 2D polycrystals performed with and without GB shear coupling to identify how shear coupling modifies essential features of microstructure evolution; modeling details are in the Methods section.
The initial polycrystalline microstructures for the grain growth simulations are based on PF simulations that are nearly equivalent to Voronoi tessellations of a Poisson point process with 1000 grains (see \cite{sal2022disconnection}).
We assume that the 2D polycrystals are composed of [110]-textured face-centered cubic grains (there are two sets of closely-packed/low-energy GB planes, perpendicular to each other, in the coincidence-site lattices).
If we choose the reference system such that $[001]\,||\,\mathbf{e}_x$, $[1\bar{1}2]\,||\,\mathbf{e}_y$, and $[110]\,||\,\mathbf{e}_z$, crystal symmetry dictates that all grain orientation angles are $\theta\in$[0,$\pi$); see the orientation field of a typical initial microstructure  in Fig.~\ref{fig_poly}a.
For simplicity, all GB energies are assumed to be the same.
Therefore,  evolution without shear coupling reduces to classical mean curvature flow.
Next, we consider the effect of GB shear coupling by assigning shear coupling factors ($\beta$) to each GB according to the GB misorientation angle~\cite{homer2013phenomenology,han2018grain}; see the Supporting Information (SI) for further details on how shear coupling factors are assigned.

\begin{figure*}[t]
    \centering
\includegraphics[width=0.99\linewidth]{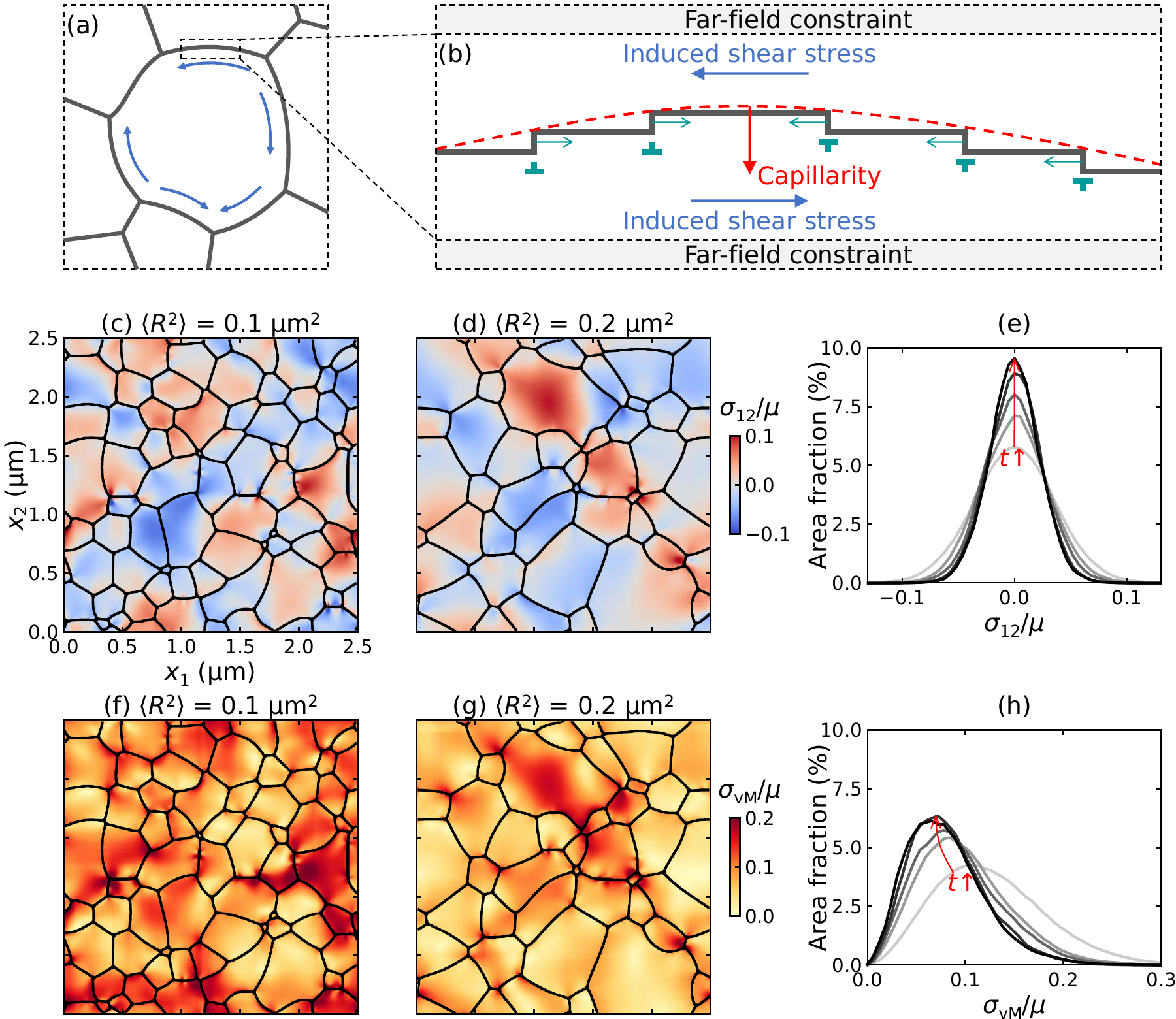}\hspace{-1.78em}%
\caption{\textbf{Internal stress fields generated by GB migration.} (a) A schematic of a grain in a polycrystal with GB shear coupling caused by disconnection motion on each GB; see the zoomed schematic in (b).
The evolving internal stress field $\sigma_{\text{12}}$ during grain growth at a grain size of (c) $1\times10^5$ and  (d) $2\times10^5$ - see Fig.~\ref{fig_poly}(d)
and (e) its distribution at different times (shear modulus is roughly that of Al).
(f)-(h) Same as (c)-(e) for the  von Mises stress.
}
\label{fig_stress}
\end{figure*}

Figure~\ref{fig_poly}b shows the evolution of the mean grain (area) size ($\langle R^2 \rangle$) with and without GB shear coupling.
Each solid, thick curve is the averaged result of 30 independent PF simulations (i.e., the light, thin curves) with different initial microstructures (i.e., different random initial microstructures and orientation distributions).
These results show that the parabolic grain growth law is approximately valid with or without shear coupling.
The grain growth rate is, however,  $\sim50\%$ larger with shear coupling than without, pointing to the importance of elastic effects.
Figures~\ref{fig_poly}c,d show the evolution of the microstructure starting from the same initial configuration, Fig.~\ref{fig_poly}a.
Shear coupling leads to significant differences in the temporal evolution of the microstructures: i.e.,  both the GB network and grain orientations.

Shear-coupled GB migration is mediated by the flow of disconnections--i.e., line defects with both step and dislocation character constrained to and lying on GBs \cite{bollmann1970general,hirth1973grain,han2018grain}.
Thus, GB shear coupling can lead to the generation of internal stress.
Figure~\ref{fig_stress}a shows a schematic of disconnection flow and the resultant shear along the GBs delineating a single grain in a polycrystal.
Focussing on the GB segment inside the dashed rectangle in Fig.~\ref{fig_stress}b, we see that  macroscopically curved GBs can be described in terms of microscopic/atomistic disconnections.
The step character is represented by the horizontal and vertical black lines, and the dislocation character of the disconnections is indicated in green.
Assuming that this GB segment is migrating downward (i.e., toward its mean curvature/capillarity), such GB migration occurs through the motion of the two vertical steps toward each other (see the green arrows).
At the same time, the two oppositely signed dislocations (Burgers vectors) glide towards each other, leading to a relative shear across the GB; this is the origin/mechanism of shear coupling.
As Burgers vector flow along each GB, a spatially and temporally varying internal stress field is produced.

Figures~\ref{fig_stress}c,d show the internal stress  $\sigma_{12}$ distribution for the  microstructures in Fig.~\ref{fig_poly}d.
The generated internal stress constantly evolves and is spatially inhomogeneous within the whole microstructure and each grain.
Figure~\ref{fig_stress}e shows the evolution of the fractions of $\sigma_{12}$; the lightest and darkest lines are for the initial and final microstructures, respectively.
The $\sigma_{12}$ stress distributions are nearly Gaussian, with widths that decrease during grain growth.
To eliminate the effect of the coordinate system and investigate the elastic energy density evolution, we also show the von Mises stresses in Figs.~\ref{fig_stress}f,g for the same microstructures.
This stress distribution is also inhomogeneous and evolving.
The von Mises stresses are nearly Rayleigh distributed, with peaks moving to lower stress during microstructure evolution (Fig.~\ref{fig_stress}h).
Both Figs.~\ref{fig_stress}e and \ref{fig_stress}h demonstrate that internal stresses are relieved by grain growth; i.e., grain growth dissipates elastic energy.

\begin{figure}[t]
    \centering
\includegraphics[width=0.95\linewidth]{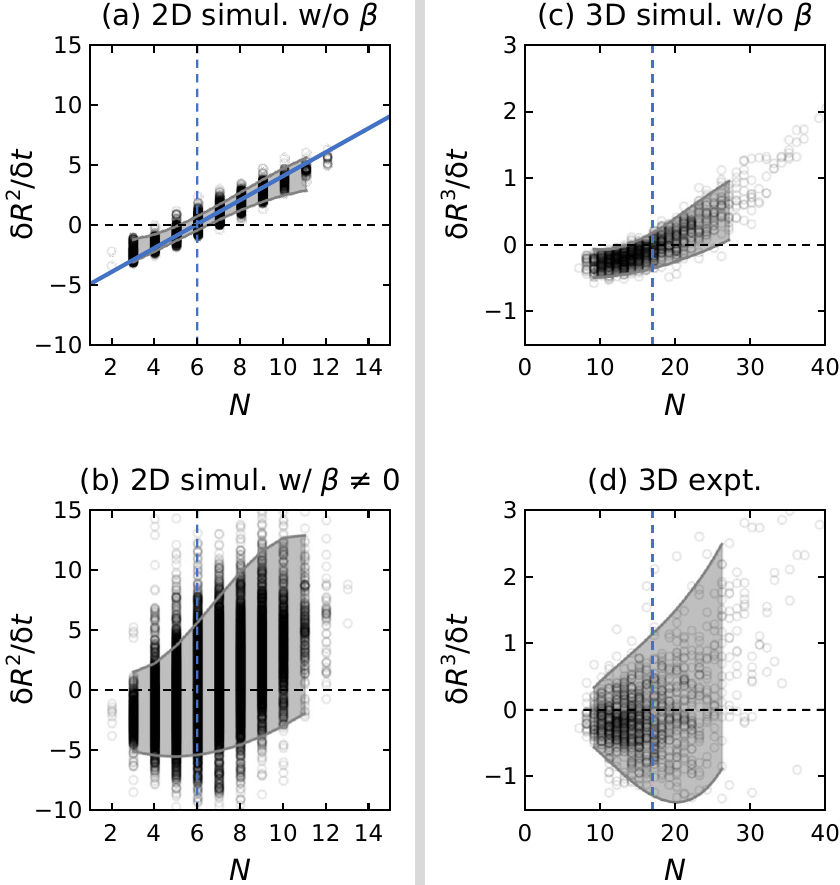}\hspace{-1.78em}%
\caption{\textbf{Failure of the von Neumann-Mullins relation.} Scatter plots of individual grain area changes $\delta R^2/\delta t$ versus the number of neighbors $N$ during grain growth  (a) without shear coupling (i.e., mean curvature flow) and (b) with shear coupling from 2D PF simulations.
Reported grain volume change $\delta R^3/\delta t$ vs. the number of neighbors in 3D from (c)  curvature flow grain growth simulations~\cite{muralikrishnan2023observations} and (d) a polycrystalline SrTiO$_3$ grain growth experiment~\cite{muralikrishnan2023observations}.
(c) and (d) are reproduced with permission from \cite{muralikrishnan2023observations} (Copyright 2023 Elsevier)
The gray shaded regions in the plots indicate the 99\% confidence intervals from the data shown.
}
\label{fig_Neumann}
\end{figure}

Note that stresses observed in the simulations shown in Fig.~\ref{fig_stress} exhibit maximum stresses as large as  $\sim\mu/10$, which are comparable to the theoretical strength of the material.
Where these large stresses are highly localized near the triple junctions (TJs), the origin of these high stresses is related to the constrained GB sliding at TJs.
We also note that there can be very large stresses within grain interiors as well.
In most metals, such large shear stress could be relieved by some plastic deformation within the grains.
While the possibility of bulk plastic deformation is not included in these simulations, these results suggest that some microplasticity (dislocation, twin) may result from GB migration in real polycrystals \cite{fullman1951formation,holm2010grain,bizana2023kinetics}.

The existence and evolution of internal stresses have also been observed in grain growth experiments \cite{hayashi2019intragranular} and MD simulations \cite{thomas2017reconciling}.
Such stresses arise naturally from shear coupling; we return to this point later.
The central question that remains to be answered is whether the mutually affected GB shear coupling and generated internal stress are not only vaguely influencing the outcome of grain growth but also allow for explaining two recent findings in pivotal experiments for polycrystals \cite{bhattacharya2021grain,xu2024grain,bizana2023kinetics}: (i) grain size change is not correlated with the number of neighbors of each grain, and (ii) GB velocity is not correlated with GB mean curvature.

\subsection{Failure of the von Neumann-Mullins relation}
In the 1950's, von Neumann \cite{von1952metal} and Mullins \cite{mullins1956two} rigorously demonstrated that if isotropic grain growth is curvature flow, the size of each grain evolves as
{\small\begin{equation}\label{eq:neumann}
    \frac{\rmd R^2}{\rmd t} = -M\gamma\left(2\pi - \sum\limits_{i=1}^{N} \alpha_i  \right) = -2\pi M\gamma \left(1-\frac{1}{6}N\right),
\end{equation}}
where $R^2$ is the grain area, $M$ and $\gamma$ are the isotropic GB mobility and  GB energy, $N$ is the number of neighbors of the grain, $\alpha_i$ is the dihedral angle at a triple junction; in the isotropic case $\alpha_i=\pi/3$.
This implies that grains with more/less than six neighbors will grow/shrink.

Figure~\ref{fig_Neumann}a shows grain area change vs. the number of neighbors of individual grains obtained in our 2D PF grain growth simulations (shown in Fig.~\ref{fig_poly})  in the absence of GB shear coupling.
The thick blue line is the ideal grain growth relation (\ref{eq:neumann}), and unsurprisingly, all data points are clustered very close to this line.
Almost all grains with less than six neighbors shrink, while those with more than six grow, consistent with the von Neumann-Mullins relation (having six neighbors implies that the mean GB curvature is zero).
While the von Neumann-Mullins grain growth relation is exact, the small discrepancy in Fig.~\ref{fig_Neumann}a is associated with the diffuse boundary nature of the simulations.
However, when the simulations include GB shear coupling, very large deviations from the ideal grain growth relation are observed; $c.f.$, Figs.~\ref{fig_Neumann}a and b.
In this case, we note that the grain growth rate data has a very large standard deviation.
This implies that the number of neighbors is no longer a good predictor of whether a grain will shrink or grow; in other words, the von Neumann-Mullins relationship fails in the presence of shear coupling.
Nonetheless, on average, grains with a large number of neighbors tend to grow, and those with few neighbors tend to shrink.

In 3D, curvature flow simulations (no shear coupling)~\cite{muralikrishnan2023observations} show that the grain volume growth rate versus number of neighbors is similar to the 2D curvature flow simulation results (i.e., the standard deviation of the data is small and there is a critical $N$ above which grains tend to grow and below which they shrink); $c.f.$, Figs.~\ref{fig_Neumann} a and c.
However, grain growth experiments on 3D SrTiO$_3$ show that the standard deviation of the grain volume growth rate is very large; see Fig.~\ref{fig_Neumann}d.
The 3D curvature flow simulation results and the 3D experiments differ because the experiments (unavoidably) include shear coupling, while the 3D curvature flow simulations do not.
This is exactly analogous to the 2D results, where simulations with and without shear coupling differ in the same manner.\footnote{Note, the extension of the von Neumann-Mullins relation to 3D shows that the grain size growth rate should be proportional to the mean width of the grain \cite{macpherson2007vonneumann}, not the number of neighbors; unfortunately, such data is not available from the experiments~\cite{muralikrishnan2023observations}.}

Based on these comparisons, we can conclude that  shear coupling leads to a profound violation of the most fundamental grain growth relation.
This strongly suggests that, in polycrystals, grain growth is not purely mean curvature flow.

\subsection{Failure of mean curvature flow}

\begin{figure*}[t]
    \centering
\includegraphics[width=0.955\linewidth]{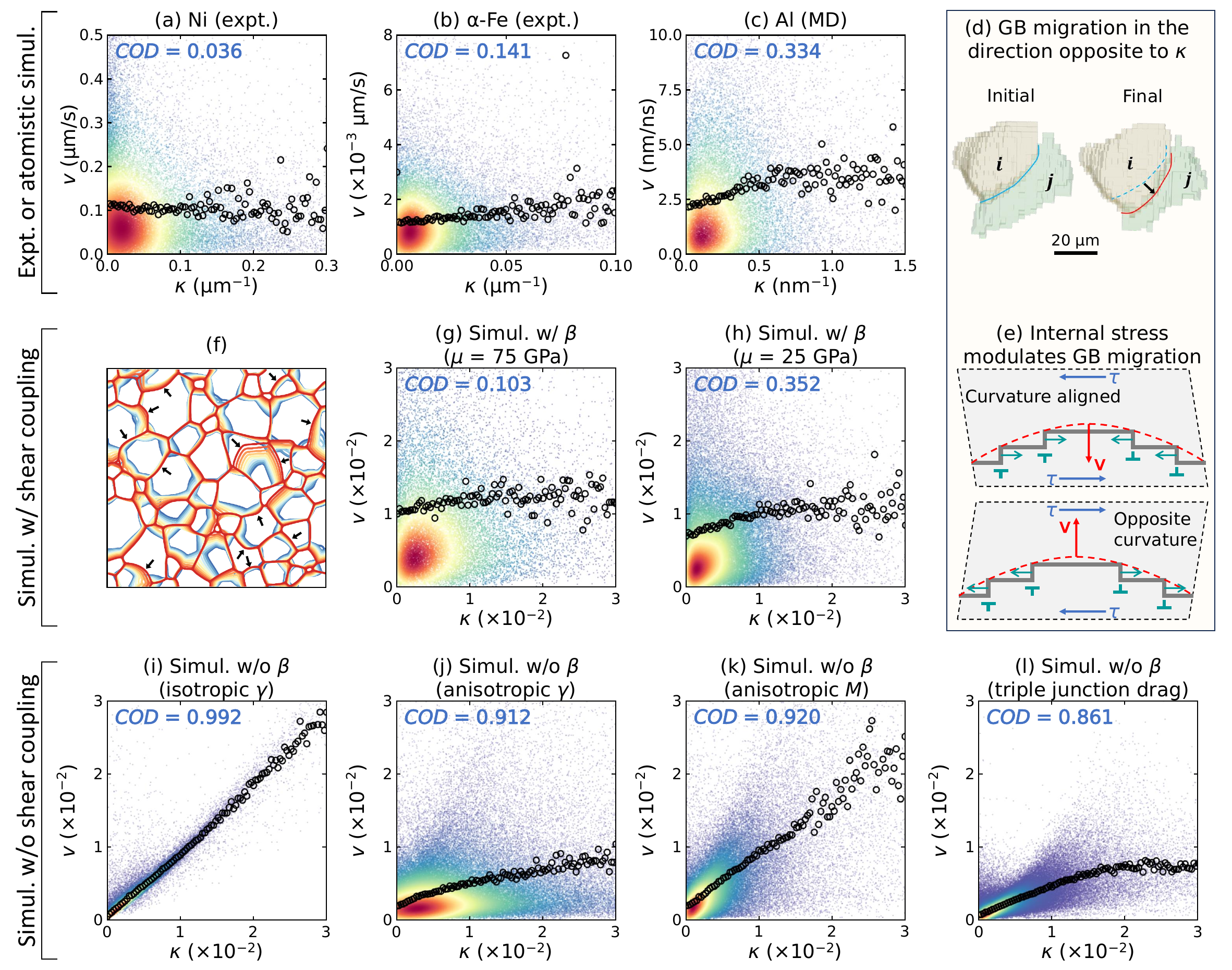}
\caption{\textbf{Failure of mean curvature flow.} Scatter plots (small  points) of GB velocity vs.  mean curvature for individual GBs in polycrystalline (a) Ni (experiment), (b) $\alpha$-Fe (experiment), and (c) Al (MD simulation). The color of each of the small points indicates the density of points (measurements) in the local vicinity of each point (redder points indicate there are many GBs with similar velocities and curvatures.
The open circles represent GB velocities averaged over GB segments with (approximately) the same mean curvature.
(d) An example of GB migration opposite to its center of mean curvature in grain growth experiments in (b).
(e) Illustration of GB migration mediated by disconnection motion in opposite directions.
(f) Simulated grain growth microstructure evolution with GB shear coupling ($\mu=25$ GPa).
Here, increasing time is indicated by changes in the GB colors from blue to red. 
Black arrows indicate locations where the GBs migrate in the sense opposite to the curvature.
(g,h,i) Same type of plots (a)-(c) for PF  grain growth simulations with shear coupling for polycrystals with shear modulus (e) $\mu = 75$ GPa, (f) $\mu = 25$ GPa and (g) with no shear coupling. $\mu = 0$.
(j)-(l) Same as (i) but with (j) anisotropic GB energy, (k) anisotropic GB mobility and (l) triple junction drag (see the text).
(a)-(d) are reproduced with permission from~\cite{bhattacharya2021grain,xu2024grain,bizana2023kinetics} (Copyright 2021 AAAS, 2023 Elsevier, 2024  Elsevier).
}
\label{fig_correlation}
\end{figure*}

The von Neumann-Mullins relation (Eq.~\eqref{eq:neumann}) stems from the mean curvature flow assumption, $v = M\gamma\kappa$, i.e., GB velocity is proportional to the mean GB curvature.
However, statistical analysis of experimental measurements on nickel \cite{bhattacharya2021grain} and iron \cite{xu2024grain} showed that GB velocity is  not (or only weakly) correlated with mean curvature (data reproduced in Figs.~\ref{fig_correlation}a and b).
Similar conclusions were drawn from MD simulations of polycrystalline  aluminum (Al) \cite{bizana2023kinetics}; see Fig.~\ref{fig_correlation}c.
The small colored dots in these figures are individual measurements of the velocity and (weighted) mean curvature of different GBs; the color of each dot indicates the density of dots (measurements) in the local vicinity of each dot. Redder dots indicate there are many GBs with similar velocities and curvatures.
The open black circles refer to the average GB velocity of all GBs with the same (weighted) mean curvature (binned data).

Clearly, any relationship between the GB velocity and mean curvature is very weak; certainly $v = M\gamma\kappa$ is inconsistent with these individual GB data.
This is shown by noting the very small correlation between the average GB velocity and mean GB curvature (see the open circles); the coefficient of determination $COD$ (i.e., the linear correlation, often called $R$ squared) for the experiments and simulations are all less than $0.35$.
Since the GB velocity and GB mean curvature are only weakly correlated, the fundamental assumption that led to the von Neumann-Mullins relation (or its extensions to higher dimensions or inclusion of anisotropic GB energy) is invalid.
Note that while the GB velocity is not proportional to GB curvature, capillarity (reduction in GB energy) remains the driving force for normal grain growth; curvature flow is necessarily modified by shear coupling.

Why is the GB velocity not (only weakly) correlated with GB mean curvature?
The simulation results above strongly suggest that this is related to shear coupling.
As noted (see Fig.~\ref{fig_stress}a-b), GB migration and shear coupling are mediated by the motion of disconnections.
A more appropriate equation of motion for GBs, that accounts for shear coupling in addition to curvature effects, can be expressed as \cite{zhang2017equation,han2022disconnection}
\begin{equation}\label{EOM}
\mathbf{v} = \mathbf{M}
\left(\Gamma\kappa + \boldsymbol{\tau} \cdot \boldsymbol{\beta} + \psi\right)
\hat{\mathbf{n}},
\end{equation}
where $\mathbf{v}$ is the  velocity of a GB segment and $\mathbf{M}$ is the intrinsic \emph{disconnection} mobility tensor.
This equation of motion features three driving forces.
$\Gamma \kappa$ is the weighted mean curvature/classical capillarity driving force, with interface stiffness $\Gamma = \gamma + \gamma_{,\phi\phi}$ ($\phi$ is the local inclination angle) and local mean curvature $\kappa$.
$\psi$ is the chemical potential jump across the GB related to differences in the bulk free energy densities of two grains meeting at the interface ($\psi = 0$ in normal grain growth).
Both $\mathbf{M}$ and $\Gamma$ retain anisotropies related to the crystal structure.
$\boldsymbol{\tau} \cdot \boldsymbol{\beta}$ represents the elastic driving force acting on the disconnection Burgers vectors related to the GB shear coupling, where $\boldsymbol{\tau}$ and $\boldsymbol{\beta}$ are the resolved internal shear stress and shear-coupling factor vectors. 
If only isotropic capillarity forces act on the GB, this equation of motion reduces to GB curvature flow.

Eq.~\eqref{EOM} implies that GBs can migrate either toward or against its curvature due to the existence of internal stresses.
Consider the situation in the schematic in Fig.~\ref{fig_correlation}e
where the capillarity force drives the GB segment downward (toward its center of curvature) via the motion of the two vertical disconnections towards each other.
However, the internal stress induced by one disconnection leads to elastic interactions in terms of a Peach-Koehler force on the Burgers vector of the other disconnection.
If the total Peach-Koehler force acting on the disconnections drives them toward each other, the GB segment will migrate in the same direction as when driven by curvature albeit faster.
On the other hand, if the Peach-Koehler force drives these two disconnections apart, the resultant driving force is opposite to that of mean curvature flow; for sufficiently large Peach-Koehler forces the GB segment will migrate in the direction opposite to that in curvature flow.
In any case, this elastic effect weakens the correlation between GB velocity and curvature.  
GB migration in the direction opposite to the center of curvature has been observed in grain growth experiments~\cite{xu2024grain, muralikrishnan2023observations}; e.g., see  Fig.~\ref{fig_correlation}d from Ref.~\cite{xu2024grain}.
In our grain growth simulations with GB shear coupling, we also observe GB motion in the direction opposite to the center of the curvature; see the GB network trajectory in Fig.~\ref{fig_correlation}f where indicated by the arrows.
(Increasing time in this  microstructure  is indicated by the transition of the GB color from  blue to red.)
Note that, while anisotropic $\mathbf{M}$ or $\Gamma$ in Eq.~\eqref{EOM} (i.e., weighted mean curvature flow) are essential for accurately describing the magnitude of the velocity, they cannot capture these observations of GB migration in the direction opposite to the sense of the curvature, as  illustrated in Figs.~\ref{fig_correlation}d-e, see also Figs.~S4c-f in SI.
Without shear coupling, such motion might still occur due to the torque term in $\Gamma$ for high-energy GBs \cite{FLOREZ2022117459}. This effect could then be observed in the early stages of the evolution of microstructures featuring randomly oriented GBs. As experiments typically show faceted GBs with orientations close to minimum-energy ones rather than random, the significant amount of GBs found to move opposite to curvature \cite{xu2024grain} cannot be explained by this effect only.

We now demonstrate that the main cause for the weak correlation between GB velocity and curvature is, indeed, the generation of internal stresses associated with GB shear coupling.
We analyze the GB velocity vs. GB mean curvature distributions obtained by our PF grain growth simulations with and without GB shear coupling, $c.f.$ Figs.~\ref{fig_correlation}g-i, where we employed the Bhattacharya et al.~\cite{bhattacharya2021grain} approach to extract GB velocity and mean curvature during grain growth.
When grain growth is mean curvature flow without GB shear coupling (Fig.~\ref{fig_correlation}i), a clear linear distribution and very strong correlation  ($COD \approx 1$) between GB velocity and mean curvature are found; this is unsurprising (note that while this is mean curvature flow, each data point corresponds to an average over an entire GB, rather than points on a GB; this explains the sparse deviation from the exact, isotropic, von Neumann-Mullins law).
Figures~\ref{fig_correlation}g and h show what happens when GB shear coupling is included; the correlation between curvature and velocity is not only lost, but the resultant distribution of curvature-velocity scatter appears \emph{very} similar to the experimental and MD results ($c.f.$, Figs.~\ref{fig_correlation}a-c and g,h).
The strong similarity is observed both for individual GBs and for the results averaged over GBs with the same curvature, whilst the $COD$ from the PF simulations are similar to those from the experimental data.
Shear coupling generates Peach-Koehler forces which are mediated by elastic fields, hence small shear modulus $\mu$ would imply decreased shear coupling effects.
We performed PF simulations with small ($\mu=25$ GPa) and large ($\mu=75$ GPa) shear moduli ($c.f.$ , Figs.~\ref{fig_correlation}g and h) and, indeed, found that reducing the shear modulus increases the correlation between GB velocity and curvature.
This is consistent with the fact that Fe and Ni have a high modulus but low $COD$, and Al has a low modulus and higher $COD$.

The results presented above provide strong, clear evidence that the (experimental and MD simulations) observations of weak correlation between GB velocity and curvature is a direct consequence of GB shear coupling.
Other possible reasons for this lack of correlation have been proposed: (i) GB replacement as a result of anisotropic GB energy \cite{xu2023energy}  
or anisotropic GB mobility \cite{ZHANG2020211}
and (ii) triple junction drag effect \cite{xu2024grain}.
To examine the effect of anisotropic GB energy, we conducted a series of PF simulations that include misorientation and inclination-dependent anisotropic GB energies \emph{without} GB shear coupling (using the GB energy function of Al from \cite{bulatov2014grain}); the resultant GB velocity vs. mean curvature distribution is shown in Fig.~\ref{fig_correlation}j.
The results are similar to the mean curvature flow simulation results (Fig.~\ref{fig_correlation}i), albeit with a slightly weaker curvature-velocity correlation.
Hence, GB energy anisotropy does lead to weaker velocity-curvature correlation but still very much higher than that found in either the experiments or MD simulation.
Note that GB energy anisotropy indeed leads to the replacement of high-energy GBs with low-energy GBs in our PF simulations (see an example in Fig.~S3 in SI), as in the experiments \cite{bhattacharya2021grain}. 
Figure~\ref{fig_correlation}k shows the results of simulations performed with isotropic GB energy but anisotropic GB mobility, where the magnitude of the GB mobility anisotropy is consistent with the experiments of Zhang et al.~\cite{ZHANG2020211}; see SI for details.
The results for the simulations with anisotropic GB mobility is very similar to those for anisotropic GB energy.

We also investigated the effects of triple junction drag on the GB velocity-curvature correlation by reducing the TJ mobility (i.e., by lowering the PF kinetic coefficient in the triple junction region by 10$\times$ \cite{johnson2014phase}).
As noted by Gottstein and Shvindlerman \cite{gottstein2000effect}, when grain growth is controlled by TJ drag, the grain growth law is completely different from the von Neumann-Mullins law.
In our PF simulations, we find that when the GB mean curvature is low, the curvature-velocity correlation is high, but when the mean curvature is large, the correlation is notably weaker (see Fig.~\ref{fig_correlation}l).
Nonetheless, including TJ drag does, indeed, lower the curvature-velocity correlation, but the $COD$ remains much larger than that found in the experiments or MD simulations.
It is also interesting to note that the individual GB velocity-curvature measurements (small colored points in Fig.~\ref{fig_correlation}l) are distributed much differently than those in the experiments or MD simulations (Figs.~\ref{fig_correlation}a-c).
Hence, we conclude that TJ drag is not the dominant cause of the failure of curvature flow.

\begin{figure*}[tb]
  \centering
\includegraphics[width=0.78\linewidth]{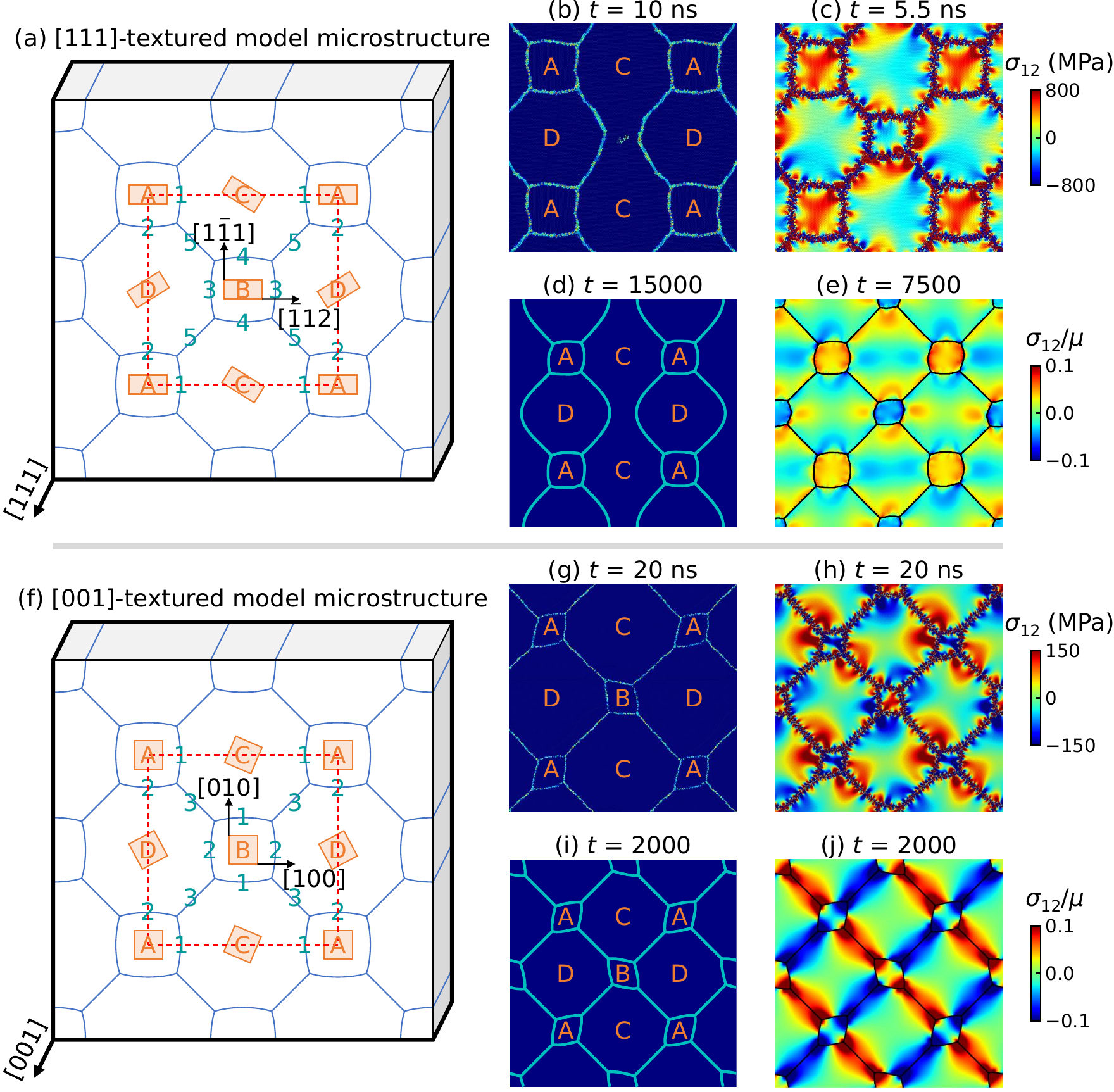}\hspace{-1.78em}%
\caption{\textbf{Model microstructures for benchmarking.} (a) Schemetic of the initial [111]-textured idealized microstructure for the reported MD and current PF grain growth simulations in (b)-(e).
Microstructure evolution in (b) reported MD simulations at time around 10 ns and (d) current PF simulations at time of 15000.
Internal stress field $\sigma_{\text{12}}$ at time (c) 5.5 ns in the MD simulations and (e) 7500 in the PF simulations.
(f) Schemetic of the initial [001]-textured idealized microstructure for the MD and PF grain growth simulations in (g)-(j).
Microstructure evolution in (g) MD simulations at time around 20 ns and (h) PF simulations at time of 2000.
Internal stress field $\sigma_{\text{12}}$ at time (i) 20 ns in the MD simulations and (j) 2000 in the PF simulations.
(b), (c) are  reproduced with permission from~\cite{thomas2017reconciling} (Copyright 2017 Springer Nature)
}
\label{fig_ideal}
\end{figure*}

While anisotropic GB energy, mobility, and TJ drag reduce the correlation between GB velocity and curvature, it is clear that shear coupling is the main source of the failure of curvature flow in grain growth.

\section{Discussion}

The effects of shear coupling on the large-scale microstructure evolution outlined above follow from the microscopic mechanism of disconnection flow. We validate this by showing that grain growth simulations based on the considered continuum (PF) model are fully consistent with the outcome of MD simulations.

We consider the idealized (small) microstructure in Fig.~\ref{fig_ideal}a, which consists of two sets of square columnar grains (A, B; same initial grain size and orientation) and two sets of octagon columnar grains (C,D; different grain orientations), and was recently examined via  MD simulations \cite{thomas2017reconciling}.
The von Neumann-Mullins relation (Eq.~\eqref{eq:neumann}) suggests that the two sets of square grains will behave similarly since they share the same GB types and number of sides.
However, the MD simulations showed that while grain B shrank, the size of grain A changed little; see Figs.~\ref{fig_ideal}b.
Figure~\ref{fig_ideal}c shows the internal stress field from the MD simulations; the stresses in grains A and B were different, implying  different GB shear coupling behaviors of the two grains.
The  stress fields in these figures show that the sources of these stresses are localized to points on the GBs (i.e., the disconnections) and the TJs (where disconnections pile-up) \cite{thomas2019disconnection}.

We conduct  PF simulations on the same idealized microstructure using material parameters determined from bicrystal MD simulations~\cite{thomas2017reconciling,homer2013phenomenology} (see the SI); i.e., there are no adjustable \emph{physical} parameters.
Without GB shear coupling, the shrinkage rates of grains A and B are, unsurprisingly, the same (see Figs.~S5a,c in SI); i.e., this microstructure evolution is not consistent with the MD simulations results.
Next, we performed PF simulations on the same microstructure, but we assigned the measured shear coupling factors (from bicrystal MD simulations) to the individual GBs.
Detailed comparisons of the PF (Figs.~\ref{fig_ideal}d,e) and MD  (Figs.~\ref{fig_ideal}b,c) microstructure simulations show that the PF simulations with shear coupling accurately reproduce the MD microstructure evolution, internal stress fields, and the difference in shrinkage rates of grain A and B.

A second example of the evolution of this idealized microstructure with different texture (tilt axis along thickness direction) and initial grain orientations results in analogous conclusions, see the schematic initial configuration in Fig.~\ref{fig_ideal}f; here too, the square columnar grains A and B share the same orientation.
In MD simulations, the shrinkage rates of grains A and B are almost the same, whilst their shapes are elongated along one diagonal and compressed along the other; see Fig.~\ref{fig_ideal}g.
We also find the different stresses in grains A and B (see Fig.~\ref{fig_ideal}h).
By assigning the shear coupling factors to individual GBs, the microstructure evolution and internal stress field obtained by our PF simulation match very well again with the MD simulation, $c.f.$ Figs.~\ref{fig_ideal}i,j and Figs.~\ref{fig_ideal}g,h.
These results  (i) demonstrate that the PF model with shear coupling provides an accurate representation of polycrystalline microstructure evolution and (ii) further demonstrate that the observation that GB velocity is not proportional to GB curvature (in the presence of shear coupling) translates into the failure of the von Neumann-Mullins law, Eq.~\eqref{eq:neumann}, at the microstructural level.
 
Increasing temperature tends to decrease the magnitude of the GB shear coupling as a result of activation of multiple disconnection modes~\cite{homer2013phenomenology,chen2019grain,han2018grain,wei2019continuum,qiu2024disconnection}.
Such a weakened shear coupling factor can improve the correlation between $v$ and $\kappa$, similar to the effect by decreasing the shear modulus as shown in Figs.~\ref{fig_correlation}g and h (shear moduli also slowly decrease with increasing temperature).
In the limit that $T\rightarrow\infty$ (ignoring melting), grain growth can be considered as curvature flow. 
The magnitude of this effect is material- (and GB-) dependent. 
Atomistic simulations show that grain growths in Al polycrystal at 0.75 $T_{\rm m}$~\cite{bizana2023kinetics} and Ni polycrystal at 0.85 $T_{\rm m}$~\cite{thomas2017reconciling} are still significantly different from curvature flow.
Hence, at most temperatures where grain growth experiments and simulations are conducted, GB shear coupling cannot be ignored to predict and understand realistic grain growth.

In summary, we examined the effects of GB shear coupling on the microstructure evolution through a series of large-scale PF simulations of grain growth in two-dimensional (2D) polycrystals.
The PF simulations are based on a bicrystallography-respecting continuum model for disconnection-mediated GB migration (see the SI for a discussion of the physical parameters in the PF simulations). 
The inclusion of shear coupling (disconnection) effects provides a clear explanation of why grain growth is not weighted mean curvature flow (these are captured in the simulations).
More specifically, shear coupling generates internal stresses that provide important additional (to capillarity) driving forces for grain boundary migration. 
These mechanical driving forces modify the classical GB equation of motion (i.e., mean curvature flow and its anisotropic extensions), thereby rendering the classical 2D von Neumann-Mullins model (and its extensions to higher dimensions and anisotropy) unable to describe grain growth.
While the effects of anisotropic GB energy and mobility do, indeed, decrease the correlation between GB velocity and weighted mean curvature, they are too weak to account for the observed deviations (both in magnitude and trends) of the experimental and simulation data from weighted mean curvature flow.
We emphasize that the ultimate driving force for grain growth remains the reduction in GB energy (capillarity), as in classical models, but the inclusion of elasticity/shear coupling is essential in the GB equation of motion in order to describe the kinetic process of grain growth.

\section{Methods}
A series of PF simulations are performed based on the diffuse interface approximation of the (sharp-interface) ~Eq.~\eqref{EOM} proposed in \cite{han2022disconnection,sal2022disconnection}.
In brief, the mean-curvature term ($\Gamma \kappa$) is approximated via a classical multi-phase field (PF) model  \cite{Chen2002,Steinbach_2009} and is coupled to an advection term encoding the other contributions.
A minor difference to the approach introduced in \cite{sal2022disconnection} is using another established PF formulation for the mean curvature flow \cite{Steinbach1999} that allows for a more versatile parametrization.

Grains forming a microstructure are described via smooth, order parameters $\{ \eta_i(\mathbf{r}) \}$ with $\mathbf{r}\in \Omega \subset \mathbb{R}^2$, which are 1 in the region corresponding to the $i$-th grain and 0 elsewhere with a smooth transition in between.
Such smooth interfaces between grains correspond to GBs, thus, in practice, corresponding to the regions where pairs of order parameters are non-zero ($\eta_i\eta_j>0$).
Note that physical parameters related to grains are labeled by one index ($i$ or $j$), while the ones of GBs are labeled by the two indexes $ij$.
The approach incorporates capillarity-driven dynamics and jumps of the energy across boundaries following the definition of the energy functional \cite{Steinbach1999}
\begin{equation}\label{eq:free-energy-PF}
\begin{split}
    F=\int_\Omega {\rm d}\Omega \bigg[ & \sum_i^N \sum_j^N \frac{4\gamma_{ij}}{\epsilon}\bigg(\eta_i\eta_j-\frac{\epsilon^2}{\pi^2}\nabla \eta_i \cdot \nabla \eta_j \bigg)\\ + & \sum_i^N \eta_i \psi_i + \lambda \bigg(\sum_i^N \eta_i - 1\bigg) \bigg],
\end{split}
\end{equation}
with $\gamma_{ij}$ the interface/grain-boundary energy density, $\epsilon$ a parameter controlling the extension of the diffuse interface between grains, $\psi_i$ the bulk free energy, and $\lambda$ a Lagrange multiplier.
The first term approximates the interface energy, and the last term is a constraint, imposing that in every point in $\Omega$, the sum of all the phase fields is 1 (i.e., enforcing complete filling of the domain with grains).
Leveraging the definition of the free energy Eq.~\eqref{eq:free-energy-PF}, the motion of GBs dictated by Eq.~\eqref{EOM} can be approximated by the following evolution law of $\eta_i$:
\begin{equation}\label{eq:eom-pf}
    \dot \eta_i=\sum_{j}^N M_{ij} \left(\frac{\delta F}{\delta \eta_j}-\frac{\delta F}{\delta \eta_i} + f_{\rm elastic}\right).
\end{equation}
with
\begin{equation}
\frac{\delta F}{\delta \eta_j}-\frac{\delta F}{\delta \eta_i}=f_{\rm curvature} + f_{\rm chemical},
\end{equation}
\begin{equation}
f_{\rm curvature} = \Gamma_{ij} \left(\eta_j\nabla^2\eta_i - \eta_i\nabla^2\eta_j-\frac{\pi^2}{2\epsilon^2}(\eta_j-\eta_i)\right),
\end{equation}
\begin{equation}
f_{\rm chemical} = \frac{\pi}{\epsilon}\sqrt{\eta_i\eta_j}(\psi_j - \psi_i),
\end{equation}
\begin{equation}
f_{\rm elastic}=\boldsymbol{\tau} \cdot \boldsymbol{\beta}_{ij} |\nabla\eta_j|(\hat{\mathbf{n}}(\eta_i) \cdot \hat{\mathbf{n}}(\eta_j))
\end{equation}
and $\hat{\mathbf{n}}(\eta)=-\nabla \eta / | \nabla \eta | $, accounting for the elastic interactions with external stress or disconnection self-stress via $\boldsymbol{\tau}$. 
This is obtained by computing the corresponding sharp-interface quantity and extending it off the interface like in Level-set approaches \cite{sal2022disconnection}. 
The continuum model of interface migration mediated by disconnection flow, Eq.~\eqref{EOM}, and thus its PF formulation, follows self-consistently from the microscopic picture of arbitrarily curved GBs with disconnections \cite{han2022disconnection,sal2022disconnection,qiu2024disconnection}. 
The density of disconnections is dictated by the geometry of the interface on which they flow, as well as dynamical/energetic effects (in a multiple-disconnection-mode formulation); see SI.
In the original formulation \cite{han2022disconnection,sal2022disconnection,qiu2024disconnection}, parameters such as GB mobility,  energy, and shear coupling factors, are expressed as a function of those for reference orientations (which define the expected disconnections). 
In this work, we set directly the shear coupling factor of the reference of arbitrarily curved GBs as a function of their misorientation \cite{han2018grain}; the corresponding full expression is reported in the SI. 
This description is shown to be fully in agreement with the microscopic picture of GB migration in the Discussion section. 
Numerical simulations are performed exploiting a simple finite-difference spatial discretization of Eq.~\eqref{eq:eom-pf} and semi-implicit time integration. Further details concerning model parametrization and additional numerical simulations are reported in the SI.

\section*{Acknowledgements}
JH acknowledges support of the National Key R \& D Program of China 2021YFA1200202 and the Early Career Scheme (ECS) grant from the Hong Kong Research Grants Council CityU21213921.
DJS acknowledges support from the Hong Kong Research Grants Council Collaborative Research Fund C1005-19G and General Research Fund 17210723.
GSR acknowledges support by the National Science Foundation under DMREF Grant No. 2118945.
MS acknowledges the support of the Emmy Noether Programme of the Deutsche Forschungsgemeinschaft (DFG, German Research Foundation) project No. 447241406.

\section*{Author Contributions}
C.Q., J.H. and M.S. designed research;
C.Q. and M.S. performed research;
C.Q. and M.S. contributed new reagents/analytic tools;
C.Q., D.J.S., G.S.R., J.H. and M.S. analyzed data;
C.Q., D.J.S., G.S.R., J.H. and M.S. wrote the paper.

\section*{Competing interests}
The authors declare no competing interests.

\clearpage
\newpage

\onecolumngrid

\begin{center}
  \textbf{\large \hspace{5pt} SUPPLEMENTARY INFORMATION \\ \vspace{0.2cm} Why Grain Growth  is Not Curvature Flow \\}
\vspace{0.4cm}   Caihao Qiu,$^{1}$, David J. Srolovitz,$^{2,3}$, Gregory S. Rohrer$^{4}$, Jian Han$^{1}$, Marco Salvalaglio$^{5,6}$ \\[.1cm]
  {\itshape \small
  ${}^1$Department of Materials Science and Engineering,\\ City University of Hong Kong,  Hong Kong SAR, China
  \\ 
  ${}^2$Department of Mechanical Engineering, The University of Hong Kong, Pokfulam Road, Hong Kong SAR, China
 \\
  ${}^3$Materials Innovation Institute for Life Sciences and Energy (MILES), The University of Hong Kong, Shenzhen, China
 \\
  ${}^4$Department of Materials Science and Engineering, Carnegie Mellon University, Pittsburgh, PA, USA
   \\
  ${}^5$Institute of Scientific Computing, TU Dresden, 01062 Dresden, Germany
   \\
  ${}^6$Dresden Center for Computational Materials Science, TU Dresden, 01062 Dresden, Germany
 }
\vspace{0.5cm}
\end{center}

\twocolumngrid

\setcounter{equation}{0}
\setcounter{figure}{0}
\setcounter{table}{0}
\setcounter{section}{0}
\setcounter{page}{1}

\renewcommand{\thesection}{S-\Roman{section}}
\renewcommand{\theequation}{S-\arabic{equation}}
\renewcommand{\thefigure}{S-\arabic{figure}}
\renewcommand{\bibnumfmt}[1]{[S#1]}
\renewcommand{\citenumfont}[1]{S#1}

\section{Diffuse interface model}
\subsection{Continuum modeling of disconnection-mediated GB motion}

The theoretical framework considered in this work builds on a continuum model for disconnection-mediated grain boundary (GB) motion that respects the underlying bicrystallography \cite{SIzhang2017equation,SIhan2022disconnection,qiu2023interface}.
We briefly summarize the continuum model as follows.
A GB is modeled as a continuous curve $\Sigma$, described by $\mathbf{x}(s)$ in the $\hat{\mathbf{e}}_1-\hat{\mathbf{e}}_2$ two-dimensional plane with $s$ a parameter.
Low energy orientation for $\Sigma$, namely low energy GBs, exist, which are considered reference orientations $R(k)$ with $k=0,\dots,N-1$ and $N$ the number of considered references.
For instance, the [110]-tilt GBs in FCC crystals possess two reference interfaces ($N = 2$), the [100]-tilt GBs exhibit four references ($N = 4$), and the [111]-tilt GBs have six references ($N = 6$).
An arbitrarily oriented $\Sigma$ may be represented as the nearest reference interfaces with a superimposed distribution of disconnections with step character $h^{(k)}$ and Burgers vector $\mathbf{b}^{(k)}$ \cite{SIhan2018grain}.
The abundance of disconnections is quantified by the respective densities $\rho^{(k)}$ and follows from geometrical considerations, namely how many disconnections per unit length are needed to obtain the actual orientation of $\Sigma$ from the closest reference orientations.
They can be thus related to the unit tangent vector $\mathbf{l}$ of $\Sigma$,
\begin{equation}\label{eq:geo}
\mathbf{l}=  \left[ {\begin{array}{cc}
   \hat{\mathbf{e}}_1 \cdot \hat{\mathbf{e}}^{(k)}  & \hat{\mathbf{e}}_1 \cdot \hat{\mathbf{e}}^{(k+1)}\\
   \hat{\mathbf{e}}_2 \cdot \hat{\mathbf{e}}^{(k)}  &  \hat{\mathbf{e}}_2 \cdot  \hat{\mathbf{e}}^{(k+1)}\\
  \end{array} } \right]
  \left[ {\begin{array}{cc}
   -h^{(k+1)}\rho^{(k+1)}  \\
   h^{(k)}\rho^{(k)}  \\
  \end{array} } \right],
\end{equation}
with $\hat{\mathbf{e}}^{(k)}$ the orientations of $k$-th reference, with $k$ and $k+1$ labelling the references with the closest orientations to $\Sigma$.
The equation of motion for the latter can be derived from the corresponding definition of the interface energy, assuming gliding of disconnections only and overdamped dynamics. It results
\begin{equation}\label{eq:eom}
    \dot{\mathbf{x}}=\mathbf{M} (\Gamma \kappa+ \boldsymbol{\tau} \cdot \boldsymbol{\beta} +\psi) \mathbf{\hat n},
\end{equation}
with $\hat{\mathbf{n}}=\mathbf{l}'(s)$ the vector normal to $\Sigma$, $\mathbf{M}$ the mobility tensor, $\Gamma=\gamma(\phi)+\gamma''(\phi)$ the interface stiffness with $\gamma(\phi)$ the energy density of the interface and $\phi$ the inclination angle, $\kappa=\nabla_s \hat{\mathbf{n}}$ the interface curvature, $\boldsymbol{\tau}=(\tau^{(k)}, -\tau^{(k+1)})^\mathsf{T}$ the resolved shear stress vector at $\mathbf{x}$, $\boldsymbol{\beta}$ the shear coupling factors, $\psi$ the jump in the chemical potential, and $\hat{\mathbf{n}}$ the vector normal to $\Sigma$.
Parameters follow from the corresponding values for references R$(k)$ and R$(k+1)$: $\mathbf{M}\equiv (M^{(k)}, M^{(k+1)})^\mathsf{T}$, $\boldsymbol{\beta}\equiv (-\beta^{(k)}, \beta^{(k+1)})^\mathsf{T}$, $\gamma(\phi) \equiv \gamma^{(k)} |\sin (\phi-\phi^{(k)})|/|\sin(\Delta \phi^{(k)})| + \gamma^{(k+1)} |\sin(\phi^{(k+1)}-\phi)|/\sin(\Delta \phi^{(k)})$, with $\Delta \phi^{(k)}= \phi^{(k+1)}- \phi^{(k)} $ (and note that $\phi^{(k)}<\phi<\phi^{(k+1)}$ by construction).

The three terms in Eq.~\eqref{eq:eom} account for capillarity-driven GB motion (first term), coupling with (shear) stress (second term), and bulk-energy differences across GBs (third term).
The shear stress includes both the effect of applied external fields and the self-stress generated by the dislocation character of all the disconnections on GBs in the system.
Explicitly, $\tau^{(k)}=(\sigma_{22}-\sigma_{11}) \sin(2\phi^{(k)})/2+\sigma_{12}\cos(2\phi^{(k)})$ with $\sigma_{ij}=\sigma_{ij}^{\rm ext}+\sigma_{ij}^{\rm self}$ ($\{i,j\}\in\{1,2\}$). $\sigma_{ij}^{\rm self}$ can be approximated as the stress field generated by a density of dislocations with Burgers vector $\mathbf{b}^{(k)}$ along the entire interface given by
\begin{equation}\label{eq:stress}
\begin{aligned}
    \sigma_{ij}^{\rm self}(s)&= K \sum_k^N \int_\Sigma |\mathbf{b}^{(k)}| \rho^{(k)}(s_0) \widetilde{\sigma}_{ij}(\mathbf{x}(s)-\mathbf{x}(s_0)) {\rm d}s_0 \\
    & = K \sum_k^N \int_\Sigma \beta^{(k)} \big[h^{(k)} \rho^{(k)}(s_0)\big] \widetilde{\sigma}_{ij}(\mathbf{x}(s)-\mathbf{x}(s_0)) {\rm d}s_0
\end{aligned}
\end{equation}
with $K=\mu/[2\pi(1-\nu)]$ and $\widetilde{\sigma}_{ij}$ are the stress field components of  a single disconnection (assuming that all disconnections only have edge dislocation character)~\cite{SIhan2022disconnection}.
Notice that we do not directly model disconnections explicitly~\cite{SIhan2022disconnection,SIsal2022disconnection,SIqiu2023interface}.
The term $h^{(k)} \rho^{(k)}$ is determined by local tangent in Eq.~\eqref{eq:geo} instead of real disconnection step heights, and the shear coupling factor effectively accounts for the Burgers vectors content.
 
The continuum model described so far applies to bicrystal configurations with one GB ($\Sigma$).
However, it can be considered for microstructures featuring a network of grain boundaries by defining parameters and geometrical properties accordingly.
This is conveniently realized in the implicit description of GBs below, allowing us also to cope easily with complex morphological evolution (typical of grain growth).

\subsection{Diffuse interface description}
To perform numerical simulations, we consider a diffuse interface approximation of the (sharp-interface) Eq.~\eqref{eq:eom} proposed in \cite{SIsal2022disconnection}.
In brief, the mean-curvature term ($\Gamma \kappa$) is approximated via a multi-phase field (PF) model  \cite{SIChen2002,SISteinbach_2009} and is coupled to an advection term encoding the other contributions.
A minor difference to the mentioned approach is using another classical PF formulation for the mean curvature flow \cite{SISteinbach1999} to allow for an extended and versatile parametrization.

Grains forming a microstructure are described via smooth, order parameters $\{ \eta_i(\mathbf{r}) \}$ with $\mathbf{r}\in \Omega \subset \mathbb{R}^2$, which are 1 in the region corresponding to the $i$-th grain and 0 elsewhere with a smooth transition in between.
Such smooth interfaces between grains correspond to GBs, thus, in practice, corresponding to the regions where pairs of order parameters are non-zero ($\eta_i\eta_j>0$).
Note that physical parameters related to grains are labeled by one index ($i$ or $j$), while the ones of GBs are labeled by the two indexes $ij$.
The approach incorporates capillarity-driven dynamics and jumps of the energy across boundaries following the definition of the energy functional \cite{SISteinbach1999}
\begin{equation}\label{eq:free-energy-PF}
\begin{split}
    F=\int_\Omega {\rm d}\Omega \bigg[ & \sum_i^N \sum_j^N \frac{4\gamma_{ij}}{\epsilon}\bigg(\eta_i\eta_j-\frac{\epsilon^2}{\pi^2}\nabla \eta_i \cdot \nabla \eta_j \bigg)\\ &+\sum_i^N \eta_i \psi_i + \lambda \bigg(\sum_i^N \eta_i - 1\bigg) \bigg],
\end{split}
\end{equation}
with $\gamma_{ij}$ the interface/grain-boundary energy, $\epsilon$ a parameter controlling the extension of the diffuse interface between grains, $\psi_i$ the bulk free energy, and $\lambda$ a Lagrange multiplier.
The double sum approximates the interface energy, and the last term is a constraint, imposing that in every point in $\Omega$, the sum of all the phase fields is 1 (i.e., enforcing complete filling of the domain with grains).
Leveraging the definition of the free energy \eqref{eq:free-energy-PF}, the motion of GBs dictated by Eq.~\eqref{eq:eom} can be approximated by the following evolution law of $\eta_i$:
\begin{equation}\label{eq:eom-pf}
    \dot \eta_i=\sum_{j}^N M_{ij} \left(\frac{\delta F}{\delta \eta_j}-\frac{\delta F}{\delta \eta_i} + f_{\rm elastic}\right),
\end{equation}
with
\begin{equation}
\frac{\delta F}{\delta \eta_j}-\frac{\delta F}{\delta \eta_i}=f_{\rm curvature} + f_{\rm chemical},
\end{equation}
\begin{equation}
f_{\rm curvature} = \Gamma_{ij} \left(\eta_j\nabla^2\eta_i - \eta_i\nabla^2\eta_j-\frac{\pi^2}{2\epsilon^2}(\eta_j-\eta_i)\right),
\end{equation}
\begin{equation}
f_{\rm chemical} = \frac{\pi}{\epsilon}\sqrt{\eta_i\eta_j}(\psi_j - \psi_i),
\end{equation}
\begin{equation}
f_{\rm elastic}=\boldsymbol{\tau} \cdot \boldsymbol{\beta}_{ij} |\nabla\eta_j|(\hat{\mathbf{n}}(\eta_i) \cdot \hat{\mathbf{n}}(\eta_j)),
\end{equation}
and $\hat{\mathbf{n}}(\eta)=-\nabla \eta / | \nabla \eta | $, accounting for the elastic interactions with external stress or disconnection self-stress via $\boldsymbol{\tau}$, obtained computing the corresponding sharp-interface quantity $\boldsymbol{\tau}$ in Eq.~\eqref{eq:stress} and extending it off the interface like in Level-set approaches \cite{SIsal2022disconnection}.

\section{Numerical Implementation and Parametrization}
\subsection{Reduced units}
For simplicity, reduced quantities are used in current PF simulations as follows:
\begin{itemize}
    \item the length quantities: $\tilde{\mathbf{x}} = \mathbf{x}/\alpha$,
    \item the time interval: $\Delta\tilde{t} = \Delta t M^{(1)}\gamma^{(1)}/\alpha^2$,
    \item the intrinsic mobility: $\tilde{M}^{(k)} = M^{(k)}/M^{(1)}$,
    \item the stress: $\tilde{\tau} = \tau\alpha/\gamma^{(1)}$,
    \item the chemical potential jump: $\tilde{\psi} = \psi\alpha/\gamma^{(1)}$,
    \item the GB energy $\tilde{\gamma} =\gamma/\gamma^{(1)}$.
\end{itemize}
$\alpha$ is chosen as 1 nm for the polycrystal PF simulations and 1 \AA~ for the idealized microstructure PF simulations.
$\gamma^{(1)}$ is set as 1 J/m$^2$ for all simulations.
Throughout the paper, we omit the tilde symbol for simplicity.

\subsection{Discretization}
Numerical simulations are performed exploiting a finite difference spatial discretization of \eqref{eq:eom-pf} and semi-implicit time integration. In practice, the domain is discretized into rectangles with edges $\rmd x_1$ and $\rmd x_2$.
In all current PF simulations, $\rmd x_1 = \rmd x_2 = \epsilon / 5$.
The discretised gradient and Laplacian operators are
\begin{equation}
    \nabla \eta(l, m) = \left[ {\begin{array}{c}
        (\eta(l+1, m) - \eta(l-1, m))/(2\rmd x_1) \\
        (\eta(l, m+1) - \eta(l, m-1))/(2\rmd x_2)\\
  \end{array} } \right],
\end{equation}
and
\begin{equation}
\begin{aligned}
    \nabla^2 \eta(l, m) = & \frac{1}{\rmd x_1\rmd x_2} [\eta(l+1, m) + \eta(l-1, m) \\ & + \eta(l, m+1) + \eta(l, m-1) - 4\eta(l, m)],
    \end{aligned}
\end{equation}
where $l$ and $m$ are the index of the element along the $\mathbf{e}_1$ and $\mathbf{e}_2$ directions.
All the simulations are performed under periodic boundary conditions in both $\mathbf{e}_1$ and $\mathbf{e}_2$ directions.

\subsection{Polycrystal}
We generate 30 samples for the PF simulations with different initial configurations.
The initial configuration for grain growth simulations consists of 1000 circle nuclei with different orientations within a background matrix.
A chemical potential difference between the nuclei and the matrix is applied so that the nuclei can rapidly consume the latter and densify into a polycrystal microstructure.
We assume that the 2D polycrystals are [110]-textured in face-centered cubic metals; i.e., two sets of closely-packed and low-energy GB planes exist and are perpendicular to each other in the CSL lattices of [110]-tilt GBs.
If we set the orientation $[001]\,||\,\mathbf{e}_x$, $[1\bar{1}2]\,||\,\mathbf{e}_y$ and $[110]\,||\,\mathbf{e}_z$ as the reference, according to crystal symmetry, the orientation angles of all grains should be in range of [0, $\pi$).

For the cases of isotropic mean curvature flow and those considering GB shear coupling (shown in Figs.~1-3 and 4f-h in main text), the GB energy ($\gamma$) and mobility ($M$) of all GBs are set as $\gamma = 1$ and $M = 4\pi^2/8\epsilon \approx 1$, where $\epsilon = 5 \rmd x$ is the interface thickness in the PF framework.
The shear coupling factor of the reference interfaces of a certain arbitrarily-curved GB is determined by their misorientation angle ($\theta$) according to \cite{SIhan2018grain}:
\begin{equation}\label{beta_microstructure}
\begin{aligned}
    &\beta^{(1)}(\theta) =
    S\left\{
        \begin{aligned}
        &2\tan\left(\dfrac{\theta}{2}\right),~ \theta < \dfrac{7\pi}{18} \\
        &2\tan\left(\dfrac{\theta - 109.53\pi/180}{2}\right),~ \dfrac{7\pi}{18} \leq \theta < \dfrac{5\pi}{6} \\
        &2\tan\left(\dfrac{\theta - \pi}{2}\right),~\dfrac{5\pi}{6} \leq \theta < \pi
        \end{aligned}
        \right. \\
         &\beta^{(2)}(\theta) = \beta^{(1)}(\pi - \theta),
    \end{aligned}
    \end{equation}
where $S$ is a scaling factor, discussed below.
The space is discretised into 250 $\times$ 250 grids with $\rmd x_1 = \rmd x_2 = 10$ (the size of the simulation domain is $L_x = L_y = 250 \times 10 \times 1$ nm = 2.5 $\mu$m), the timestep for each integration is $\Delta t = 10$.

We note that shear coupling factors measured in polycrystals tend to be significantly smaller than those found in bicrystals~\cite{SIrajabzadeh2013evidence,SIgautier2021shear,SIhomer2013phenomenology,SIhan2018grain}.
We account for this difference by introducing a shear coupling  scaling factor, $S=1/30$.
While the value of $S$ is chosen arbitrarily, it is of the correct order of magnitude compared with experimental observations.
Nonetheless, the shear coupling factor expression, Eq.~\eqref{beta_microstructure}, retains the dependence of $\beta$ on bicrystallography  consistent with theory, MD simulations, and experiments.
The source of the smaller shear coupling in polycrystals compared with bicrystals is associated with the constraints imposed by other grains (i.e., triple junctions) in a polycrystal that activate additional  disconnection modes \cite{SIthomas2017reconciling,SItian2024grain,SIcombe2021multiple}.

\subsection{Idealized Microstructure}
\label{sec:idealpar}

The initial configuration is composed of square grains A and B and octagon grains C and D, see Figs. 5a,b in the main text.
Grains A and B share the same orientation--i.e.,
$[2\bar{1}\bar{1}]\,||\,\mathbf{e}_x$, $[01\bar{1}]\,||\,\mathbf{e}_y$ and $[111]\,||\,\mathbf{e}_z$.
Grain C has a misorientation angle 27.8$^\circ$ away from grains A and B.
Grain D has a misorientation angle -32.2$^\circ$ away from grains A and B.
In this sense, the GBs between grains A and C, B and C are $\Sigma13[111]$ asymmetric tilt GBs, whilst the GBs between grains A and D, B and D are $\Sigma39[111]$ asymmetric tilt GBs.
According to the similar GB energy of all the GBs measured in \cite{SIthomas2017reconciling}, we assign the reduced GB energy $\gamma = 1$ and reduced GB mobility $M = 1$ to all GBs for simplicity.

Shear coupling factors of the reference interfaces of GBs between grains A and C, B and C are $\beta^{(1)}=\beta^{(2)}=0.05$, whilst those of the reference interfaces of GBs between grains A and D, B and D are $\beta^{(1)}=\beta^{(2)}=0.058$ \cite{SIhomer2013phenomenology,SIthomas2017reconciling}.

As for the [001]-textured idealized microstructure in copper, the initial configuration is also composed of square grains A and B and octagon grains C and D, with grains A and B sharing the same orientation, i.e.,
$[100]\,||\,\mathbf{e}_x$, $[010]\,||\,\mathbf{e}_y$ and $[001]\,||\,\mathbf{e}_z$.
Grain C has a misorientation angle of -22.6$^\circ$ away from grains A and B.
Grain D has a misorientation angle of 28.1$^\circ$ away from grains A and B.
In this sense, the GBs between grains A and C, B and C are $\Sigma13[100]$ asymmetric tilt GBs, whilst the GBs between grains A and D, B and D are $\Sigma17[100]$ asymmetric tilt GBs.
MD simulations are conducted with the EAM copper interatomic potential~\cite{SImishin2001structural} and the same simulation methodology mentioned in~\cite{SIthomas2017reconciling}.
In the PF simulation, the shear coupling factors are also assigned to the individual GBs based on their misorientations ~\cite{SIcahn2006coupling,SIhomer2013phenomenology}.

The space is discretised into 500 $\times$ 500 grids with $\rmd x_1 = \rmd x_2 = 1$, the timestep for each integration is $\Delta t = 0.1$.

\subsection{Anisotropic GB energy}
\label{sec:aniso}

In Figs.~4h,i of the main text, we introduce the misorientation-dependent GB energy.
The formulation of GB energy follows the generalized GB energy functional proposed by Bulatov, Reed, and Kumar \cite{SIbulatov2014grain}.{
\small
\begin{equation}\label{GB_E}
\begin{aligned}
    &\gamma^{(1)}(\theta) = \sin\left(\frac{\pi}{2}\frac{\theta - \theta_{\min}}{\theta_{\max} - \theta_{\min}} \right) \left( 1- a \log\sin\left(\frac{\pi}{2}\frac{\theta - \theta_{\min}}{\theta_{\max} - \theta_{\min}} \right)\right),
      \\   &\gamma^{(2)}(\theta) = \gamma^{(1)}(\pi - \theta), \\
      & \gamma(\theta, \phi) = \gamma^{(1)} |\cos\phi| +  \gamma^{(2)} |\sin\phi|
    \end{aligned}
    \end{equation}
}
where it ranges from 0 to 1 on segment $[\theta_{\min}, \theta_{\max}]$.
At the point $\theta = \theta_{\min}$, the slope is infinite, whilst the slope is zero at the point $\theta = \theta_{\max}$.
The detailed parameters in Eq.~\eqref{GB_E} are chosen for  Al \cite{SIbulatov2014grain}.

\subsection{Triple junction mobility}
\label{eq:tjdrag}

\begin{figure}[t]
\includegraphics[width=0.95\linewidth]{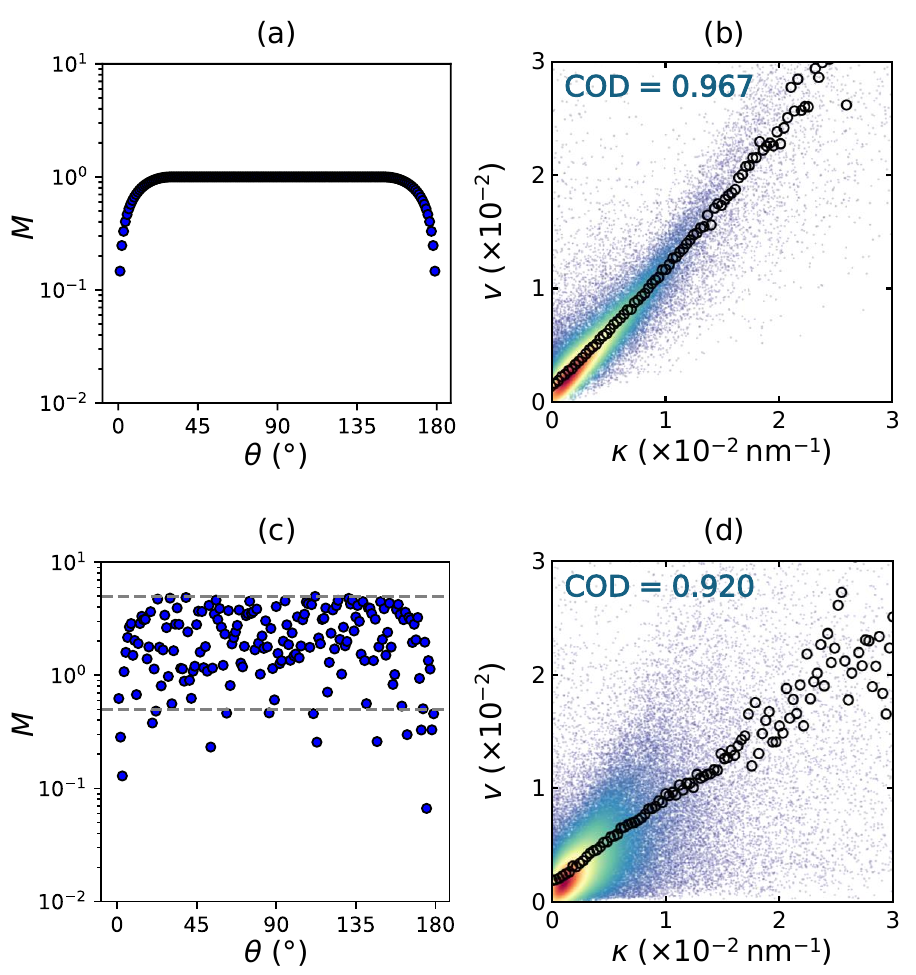}\hspace{-1.78em}%
\caption{Impact of anisotropic mobility on the correlation between velocity and GB mean curvature: (a) Misorientation-dependent GB mobility from Eq.~\eqref{eq:anisomob} and (c) (a) Randomly perturbed GB mobility resembling that reported in Ref.~\cite{SIZHANG2020211}. 
(b, d) Scatter plots (small points) of GB velocity vs. mean curvature for individual GBs obtained in grain growth considering anisotropic GB mobility in (a, c), respectively. 
The color of each small point represents the local density of measurements, with redder points indicating a higher concentration of grain boundaries (GBs) exhibiting similar velocities and curvatures.
The open circles represent GB velocities averaged over GB segments with (approximately) the same mean curvature. Note that (d) is the same as Fig. 4(l) in the main text.
}
\label{fig_mobility}
\end{figure}

\begin{figure}[t]
\includegraphics[width=0.95\linewidth]{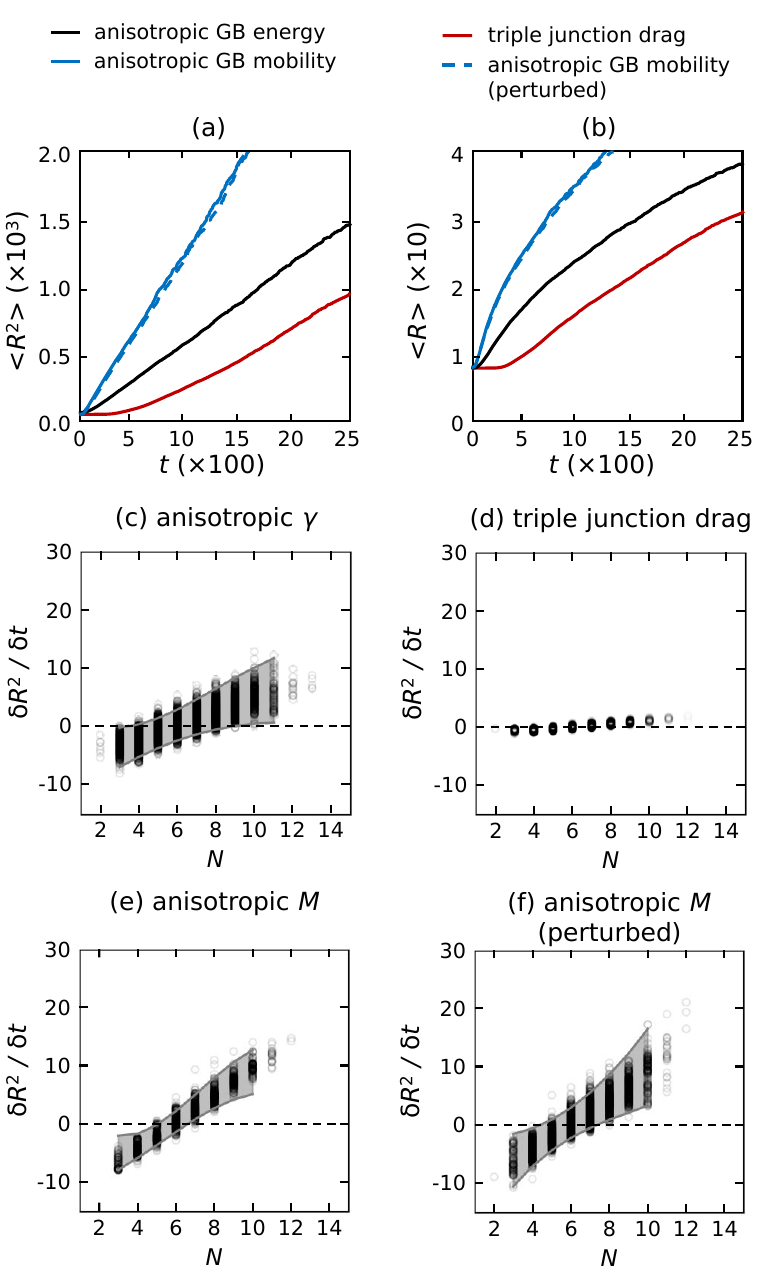}\hspace{-1.78em}%
\caption{Results on anisotropic grain growth considering anisotropic GB energy, mobility and triple junction drag effect: (a) Mean grain size $\langle R^2 \rangle$ vs. time; (b) Mean grain size $\langle R \rangle$ vs. time. Scatter plots of grain area change $\delta R^2/\delta t$ vs. number of neighbours $N$ of grains during grain growth considering: (c) anisotropic GB energy; (d) additional smaller triple junction mobility; (e) anisotropic GB mobility; and (f) perturbed anisotropic GB mobility.
}
\label{fig_anisotropic}
\end{figure}

\begin{figure*}[t]
\includegraphics[width=0.95\linewidth]{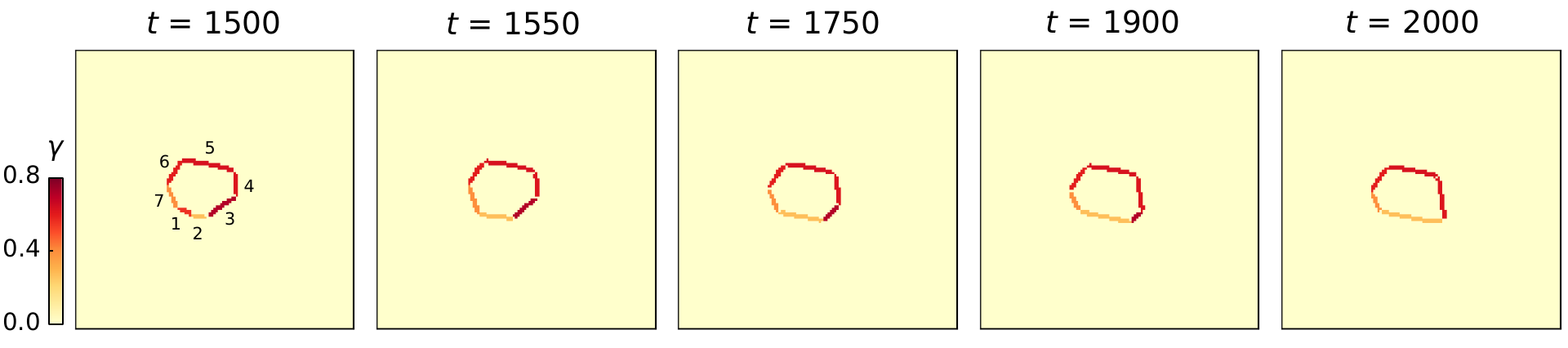}\hspace{-1.78em}%
\caption{Illustration of the GB replacement happening in one typical grain. The GBs are colored based on their GB energy.
}
\label{fig_replacement}
\end{figure*}

The triple junction (TJ) drag effect is incorporated by assigning a lower phase-field (PF) kinetic coefficient in the TJ region (i.e., lower TJ mobility) compared to the grain boundary (GB) region (i.e., GB mobility) \cite{SIjohnson2014phase}. If a grid point is located inside the grain $i$ or inside the grain $j$ or at a GB between them, the condition $\eta_i(l, m) + \eta_j (l, m) = 1$ holds. At TJs, instead, $\eta_i(l, m) + \eta_j(l, m) < 1$. The mobility $M_{ij}$ at the corresponding grid point ($l, m$) in Eq.~\eqref{eq:eom-pf} is thus set to
{\small
\begin{equation}
    M_{ij} = M_{ij}^{\rm GB} \big[M_{ij}^{\rm TJ} + \exp(-500( \eta_i + \eta_j - 1)^2 ) (1 - M_{ij}^{\rm TJ})\big].
\end{equation}
}
If $\eta_i + \eta_j = 1$ (at GBs), $M_{ij} = M_{ij}^{\rm GB}$ while if $\eta_i + \eta_j \ll 1$ (at TJs), $M_{ij} = M^{\rm GB}M_{ij}^{\rm TJ}$. In this work, we choose $M^{\rm TJ} = 0.1$.

\subsection{Anisotropic GB mobility}
We consider two cases:
\begin{enumerate}
    \item[(1)] A misorientation- and inclination-dependent GB mobility reading
        \begin{equation}\label{eq:anisomob}
        \begin{aligned}
            &M(\theta) = \left\{
                \begin{aligned}
                & M_0 \left[\frac{\theta}{\theta_{\rm c}} \left(1-\ln\frac{\theta}{\theta_{\rm c}}\right)\right],~~~~~\theta<\theta_{\rm c} \\
                & M_0,\quad\quad\quad\quad\quad\quad\quad\quad\quad~ \theta\geq\theta_{\rm c}
                \end{aligned}
                \right.,\\
            &M(\theta, \phi) = M(\theta)\cos^2\phi + M(\pi-\theta)\sin^2\phi,
        \end{aligned}
        \end{equation}
        where $\theta_{\rm c}$ is the critical misorientation angle and $M_0$ is the GB mobility of high-angle GBs (we choose $\theta_{\rm c} = \pi/6$ and $M_0 = 1$).
        The misorientation-dependent GB mobility is plotted in Fig.~\ref{fig_mobility}a.
    \item[(2)] As shown in Ref.~\cite{SIZHANG2020211}, misorientation-dependent GB mobility seems to be randomly dispersing around a constant value.
    Therefore, we here introduce a random perturbation to the GB mobility function with the values varying by one order of magnitude (with the similar characteristics to Figs. 4a and 6a in Ref.~\cite{SIZHANG2020211}, see Fig.~\ref{fig_mobility}c.
\end{enumerate}

\section{Supplementary Results and discussion}
\subsection{Anisotropic grain growth}
In the main text, we mention that we conduct PF simulations of grain growth without GB shear coupling but with anisotropic GB energy, mobility, and triple junction (TJ) drag effect. Here, we provide more results on anisotropic grain growth in Fig.~\ref{fig_mobility} and Fig.~\ref{fig_anisotropic}.

Figures~\ref{fig_mobility}b,d provide the scatter plots of GB velocity vs. mean curvature for individual GBs obtained in grain growth considering anisotropic GB mobility.  
Figure~\ref{fig_mobility}d reports the plot shown in Fig. 4h in the main text. In both cases, we show that anisotropic GB mobility can weaken the correlation between GB velocity and curvature, whilst its effect is still too weak to explain the loss of correlation observed in experiments and atomistic simulation fully illustrated in Fig. 4 in the main text.

Figures~\ref{fig_anisotropic}a and b give the grain growth law of anisotropic grain growth with and without the TJ drag effect.
The parabolic grain growth law is generally still obtained in the case of anisotropic grain growth.
With additional TJ drag (i.e., smaller TJ mobility, see \ref{eq:tjdrag}), the grain growth exponent $n$ ($R^n - R_0^n = kt$) reduces from 2 (i.e., parabolic law) to 1.32.
When the grain growth exponent $n = 1$, the grain growth is controlled by TJ motion.
Therefore, with the current TJ mobility (one order of magnitude smaller than the GB mobility), TJ effect is already clear.

Figures~\ref{fig_anisotropic}c-f show the scatter plots of grain size changes vs. the number of neighbours.
Although the grain size changes vs. the number of neighbours relation has a larger dispersion than that obtained in isotropic mean curvature flow, the data distributions (and the shapes of the upper and lower limit) are still largely different from those in experiments.
However, clear linear distributions are found, and the only difference is the reduction of the corresponding slope.

Previous studies \cite{SIxu2023energy, xu2024grain} attributed the lack of correlation between grain boundary (GB) velocity and mean curvature in polycrystalline Ni and $\alpha$-Fe 
to GB replacement driven by GB energy anisotropy. However, we demonstrate in this work that GB velocity remains correlated with mean curvature, even when accounting for GB energy anisotropy and including the GB replacement mechanism as illustrated in Fig.~\ref{fig_replacement}. Therein, high-energy boundaries 1 and 3 disappear during evolution, while the low-energy boundaries 2 and 7 increase in length.
The grain size increases less than 1\% during this procedure.
This phenomenon can be described as "GB replacement" according to the definition in \cite{SIxu2023energy, SIxu2024grain}. 

The overall results reported in this section further demonstrate that, although the phenomenology discussed in the literature, like GB replacement, is captured in our simulations, anisotropic GB energy, mobility, and TJ drag effect are still not the reason for the experimentally observed failure of the curvature flow description of grain growth.

\subsection{GB migration opposite to the mean curvature}

\begin{figure}[h!]
\includegraphics[width=0.95\linewidth]{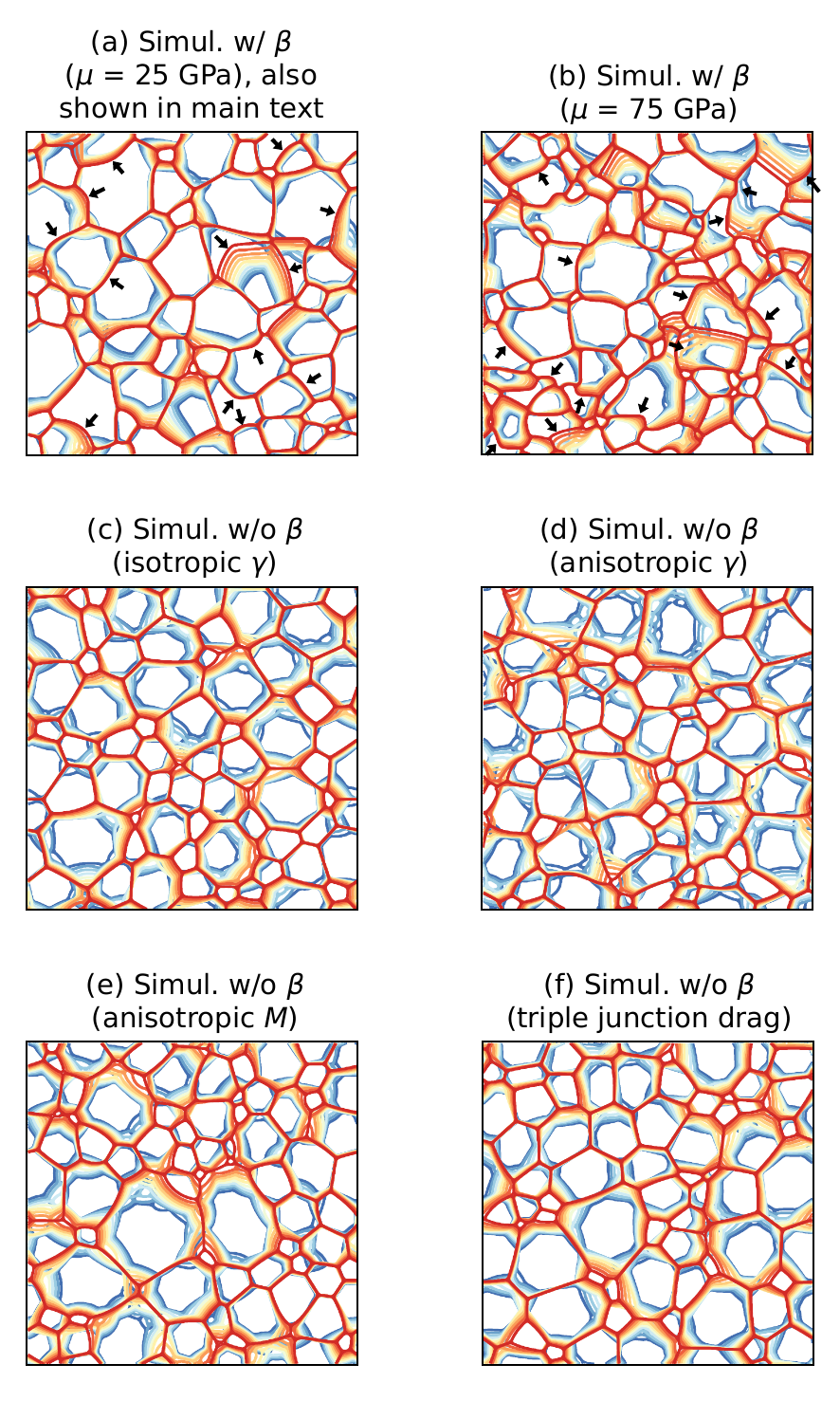}\hspace{-1.78em}%
\caption{Evolution of the GB networks during PF simulations of grain growth. (a) and (b): PF simulations including GB shear coupling with a shear modulus of 25 GPa and 75 GPa. Scenarios without shear coupling include: (c) isostropic mean curvature flow; (d) anisotropic GB energy; (e) anisotropic GB mobility; (f) triple junction drag.
With increasing time, the microstructure evolves from the blue GB networks to red GB networks.
Black arrows in (a) and (b) indicate the GBs undergoing anti-curvature GB migration. Note that panel (a) is the same as Fig. 4f in the main text.
}
\label{fig_anti}
\end{figure}

In Fig. 4f in the main text, we only provide the evolution of the GB network during grain growth simulations with GB shear coupling with shear modulus $\mu = 25$ GPa.
We show in Fig.~\ref{fig_anti} the evolutions of the GB networks in other cases. With increasing time, the microstructure evolves from the blue GB networks to red GB networks. The black arrows identify a few GBs in which the migration occurs in the direction opposite to the mean curvature. It emerges that GB migration in the sense opposite to the mean curvature can only happen in grain growth when considering GB shear coupling (Figs.~\ref{fig_anti}a,b), whilst no such ``anti-curvature'' GB migration is found in curvature flow-based grain growth simulations (Figs.~\ref{fig_anti}c-f). 

\newpage
\subsection{Evolution of the idealized microstructure \\ without shear coupling}

\begin{figure*}[t]
\includegraphics[width=0.99\linewidth]{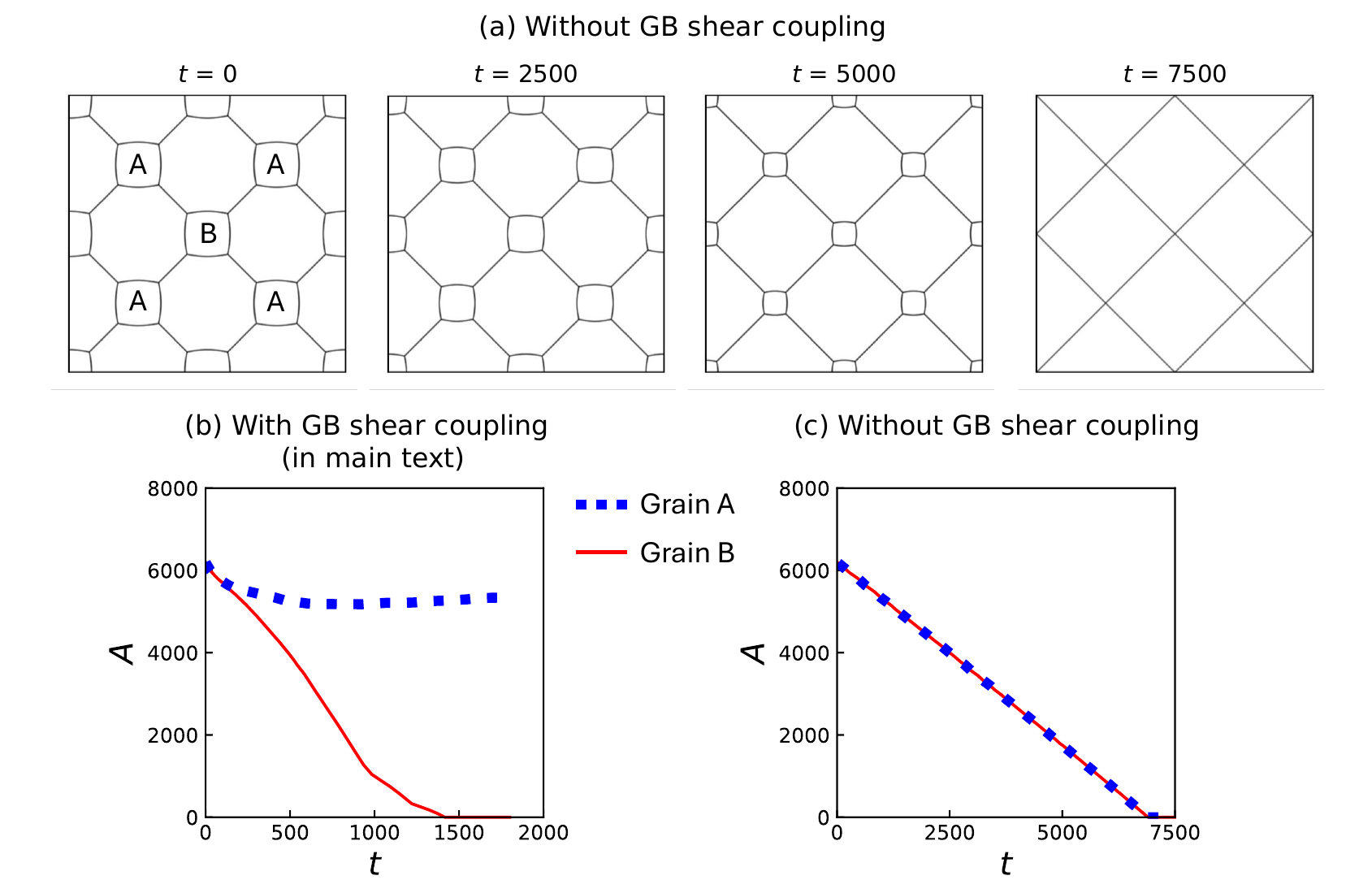}\hspace{-1.78em}%
\caption{Idealized microstructure. (a) Evolution of the [111]-texture idealized microstructure without GB shear coupling.
Grain shrinkage rates of grains A and B (b) considering GB shear coupling and (c) without GB shear coupling.
}
\label{fig_iso}
\end{figure*}

In the discussion section of the main text, an idealized (small) microstructure parametrized as detailed in Sect.~\ref{sec:idealpar} is studied. In the main text, we report the comparison between PF simulations, including shear coupling and the results of MD simulations, which show good agreement. PF simulations without shear coupling show an evolution that deviates from these MD results. In this case, as shown in Fig.~\ref{fig_iso}a, grain A and B shrink at the same rate, while a different rate is obtained including shear coupling. The evolution of the area over time of such grains with and without shear coupling is further illustrated in Fig.~\ref{fig_iso}b and c, respectively. 

\end{document}